*Article*

# DNA geometry around highly transcribed bacterial genes

**Marc Joyeux [1]**

[1] Laboratoire Interdisciplinaire de Physique, CNRS and Université Grenoble Alpes, 38400 Grenoble, France; marc.joyeux@univ-grenoble-alpes.fr

**Abstract**

Recent high-resolution (≈500 bp) Hi-C experiments reported contact maps with unusual patterns around highly expressed bacterial DNA loci. These patterns, described as “arched stripes” and “bundled domains”, extend over nearly 100 kbp. We performed Brownian Dynamics simulations with a specially designed coarse-grained model to rationalize these findings. The main feature of the model is that it takes explicitly into account the waves of positive (respectively, negative) supercoiling generated by the translocating polymerase downstream (respectively, upstream) of its position. Contact maps computed from the simulations also display the arched stripe and bundled domain patterns for above-threshold values of the rate of twist injection. Computed DNA conformations indicate that these patterns reflect an extraordinarily entangled DNA geometry, which involves both plectonemic and toroidal supercoiling, with some DNA segments experiencing both of them simultaneously. Moreover, DNA segments located on each side of the polymerase systematically wind around tracts located on the other side, which is a purely out-of-equilibrium effect driven by the continuous injection of twist. The present work therefore reveals that repeated transcription of a DNA locus systematically brings into contact DNA segments separated by several tens of kbp, which may eventually contribute to allosteric modulation and long-range communication.

**Keywords:** Gene transcription; DNA geometry; Coarse-grained model; Brownian Dynamics simulations; Hi-C contact map;

## 1. Introduction

Transcription is a mechanically and topologically active process that continuously perturbs the physical state of the DNA template. Because the RNA polymerase must progress along and transiently open a double-stranded helical DNA molecule, its movement generates torsional stress, resulting in the accumulation of positive supercoiling ahead of the polymerase and negative supercoiling behind it [1-3]. These transcription-induced changes in DNA topology can be described in terms of variations in DNA twist and writhe and, when torsional stress is constrained, can propagate over genomic distances [4]. The Hi-C methodology lends itself as a powerful tool to investigate such geometrical changes. It involves a sequence of formaldehyde cross-linking, restriction enzyme digestion, proximity ligation, and high-throughput sequencing to measure the average frequency at which DNA segments present in a cell come into close contact to each other, thereby revealing the three-dimensional organization of the genome [5]. Originally, the resolution of the technique was limited to about one Mbp, which however proved sufficient to uncover the fractal globule organization of chroma-tin [5]. Within a few years, a series of modifications were brought to the original protocol, which increased the resolution of contact maps up to one kbp [6,7]. In particular, the so-called in-situ Hi-C methodology allowed the identification of thousands of chromatin loops in the human genome [6]. More recent methodological advances contributed to further increase sensitivity and resolution for complex genomes [8,9]. Although Hi-C was originally developed to scrutinize eukaryotic species, it was soon adapted to prokaryotic ones [10-15], for which resolution below one kbp is nowadays also achieved [13-15]. Experiments performed at such high resolution reveal extraordinarily precise details of the organization of the DNA molecule. The present work precisely focuses on an intriguing high-resolution Hi-C contact map reported in [15]. In order to understand how the repeated transcription of a highly expressed locus impacts the geometry of bacterial DNA, the authors of [15] inserted a T7 promoter in the genome of *Escherichia coli* cells. They next introduced polymerase enzymes specific to the T7 promoter, while simultaneously treating the cells with rifampicin, an antibiotic which binds to bacterial RNA polymerase and blocks endogenous RNA synthesis. This led to repeated transcription starting at the phage promoter, with all endogenous genes being silenced. The resulting contact map at 500 bp resolution, extracted from Figure 2(b) of [15], is shown in Figure 1. The authors of [15] note that the contact map displays a rather unusual feature, namely the "*arched stripe*" pattern labeled (1) in Figure 1, which had hitherto not been reported. They were able to determine that the T7 promoter is probably "*positioned in the middle of the arched stripe*" [15]. Owing to the fact that a transcribing polymerase generates waves of positive supercoiling downstream of its position and negative supercoiling upstream of its position [1-3,16], they conjectured that the arched stripe may be "*the Hi-C signature of a negative supercoiled structure positioned in the upstream 5′ region of the transcription unit*" [15], a hypothesis that was supported by complementary experiments. The authors of [15] furthermore point out that the contact map also displays a thick *bundled domain* (labeled (2) in Figure 1), which "*decreases smoothly along a ~110-kb track*" [15], and depends, like the arched stripe, on transcription but not on translation. For the sake of completeness, less us finally mention that the present study uncovers a third unusual but conserved domain in contact maps, namely a *counter-arch* that opposes the arched stripe, which is labeled (3) in Figure 1.

In the present work, we used coarse-grained (CG) modeling and Brownian Dynamics (BD) simulations to go beyond the conclusions in [15] and understand in full detail the geometry of DNA molecules around highly expressed loci. DNA was coarse-grained at the level of one bead per 7.5 bp, which gives access to the dynamics of molecules that are tens of kbp long over time scales up to 100 ms. Moreover, the 7.5 bp graining is still sufficiently small with respect to the persistence length of DNA (about 150 bp) and its characteristic writhe length (about 250 bp per plectonemic turn at a superhelical density of -6%) to ensure a correct description of its geometry. This CG model was recently proposed to understand how the waves of supercoiling generated by a translocating polymerase [1-3,16] impact the DNA molecule and particularly the dynamics of its plectonemes [17,18]. However, the work in [17,18] dealt with a somewhat too short DNA molecule and did not involve the repeated transcription of a given locus. This preliminary work was therefore extended by launching BD simulations for a longer DNA molecule and a polymerase that translocates repeatedly along the same track. Computed contact maps were first compared to experimental ones to validate the CG model. Detailed analysis of the results then enabled to establish a one-to-one correspondence between the unusual features of the contact maps and the extraordinarily entangled geometry of the DNA molecule around highly expressed loci. Indeed, conformation of the DNA coil around the transcribed locus involves both plectonemic and toroidal supercoiling, with at least one long DNA segment experiencing both types of supercoiling simultaneously. Moreover, DNA segments located on both sides of the polymerase systematically wind around segments located on the other side, which is a purely out-of-equilibrium effect driven by the continuous injection of twist on both sides of the polymerase. The present work therefore reveals how the repeated transcription of a particular DNA locus systematically brings into contact DNA segments located on both sides of the polymerase and separated by as far as several tens of kbp, which may be of crucial importance, for example for allosteric modulation and long-range communication. The role of several parameters, like the superhelical density, the rate of twist injection and the length of the transcription unit was also scrutinized.

## 2. Materials and Methods

Genomic DNA is modeled as a circular chain of $n = 5760$ beads of radius $a = 1.0$ nm separated at equilibrium by a distance $l_0 = 2.5$ nm. Each bead represents 7.5 bp, so that the chain represents a DNA molecule with 43200 bp. Associated to each bead $k$ are a vector $\mathbf{r}_k$, which describes the position of its center in the space-fixed frame, and a body-fixed orthogonal frame of unit vectors $(\mathbf{f}_k, \mathbf{v}_k, \mathbf{u}_k)$, where $\mathbf{u}_k$ points from bead $k$ to bead $k+1$. $(\alpha_k, \beta_k, \gamma_k)$ denotes the set of Euler angles, which transforms $(\mathbf{f}_k, \mathbf{v}_k, \mathbf{u}_k)$ into $(\mathbf{f}_{k+1}, \mathbf{v}_{k+1}, \mathbf{u}_{k+1})$. $\beta_k$ measures the angle formed by beads $k$, $k+1$ and $k+2$, and $\alpha_k + \gamma_k$ the rotation of $(\mathbf{f}_{k+1}, \mathbf{v}_{k+1})$ with respect to $(\mathbf{f}_k, \mathbf{v}_k)$ [19]. The DNA chain is enclosed in a confinement sphere of radius $R_0 = 120$ nm, which corresponds to a concentration of base pairs around 10 mM (or 6 millions base pairs per μm$^3$), close to physiological values.

The internal energy of the DNA chain is expressed as the sum of 4 terms

$$V_{\mathrm{DNA}} = \frac{h}{2}\sum_k (l_k - l_0)^2 + \frac{g}{2}\sum_k \beta_k^2 + \frac{\tau}{2}\sum_k (\alpha_k + \gamma_k)^2 + q^2 \sum_k \sum_{m>k+3} H(|\mathbf{r}_m - \mathbf{r}_k| - 2a), \tag{1}$$

which describe the stretching, bending, torsion, and electrostatic energy of the DNA chain, respectively. The stretching term has no strict biological counterpart and was introduced to avoid a (more complex) rigid rod description of the DNA chain. $l_k = |\mathbf{r}_{k+1} - \mathbf{r}_k|$ denotes the distance between the centers of two successive beads and the stretching force constant was set to $h = 100 k_B T / l_0^2$, where $T = 298$ K is the temperature of the system, in order that the average value of $|l_k - l_0|$ does not increase beyond $l_0/10$. The bending force constant was set to $g = (k_\mathrm{B} T L_\mathrm{p})/l_0 = 20 k_\mathrm{B} T$, in order that the model reproduce the known persistence length of DNA under physiological condition, $L_\mathrm{p} \approx 50$ nm [20]. Similarly, the torsion force constant was set to $\tau = 25 k_\mathrm{B} T$, in order that the model reproduce the observed ratio of writhe to linking number difference under standard conditions, $\langle W\mathrm{r} \rangle / (L\mathrm{k} - L\mathrm{k}_0) \approx 0.7$ [21,22]. Finally, the electrostatic energy is borrowed from Model II of [23] and expressed as a sum of repulsive Debye-Hückel terms with hard core. Function $H(r)$ is defined according to

$$H(r) = \frac{1}{4\pi\varepsilon r} \exp\left(-\frac{r}{r_\mathrm{D}}\right), \tag{2}$$

where $\varepsilon = 80\varepsilon_0$ is the dielectric constant of the buffer and $r_\mathrm{D} = 1.07$ nm the Debye length in the cell. This value of the Debye length corresponds to a concentration of monovalent salt of 100 mM, which is the value that is generally assumed for the cytoplasm of bacteria. $q = -3.52\bar{e}$, where $\bar{e}$ is the absolute value of the charge of the electron, is the value of the electric charge placed at the center of each DNA bead. This value was deduced from Manning's counterion condensation theory [24,25]. Together with the small value of $l_0$ and the large value of $h$, the repulsive Debye-Hückel terms with hard core ensure that two DNA strands cannot cross each other, which is of utmost importance for the present work.

The repulsion exerted by the confinement chamber is expressed in the form

$$V_{\mathrm{wall}} = \zeta \sum_k \left[\left(1 + \frac{|\mathbf{r}_k|}{R_0}\right)^6 - 1\right], \tag{3}$$

where $\zeta = 1000k_\mathrm{B}T$, and the sum extends only to particles $k$ that satisfy $|\mathbf{r}_k| > R_0$.

The dynamics of the system was investigated by numerically integrating overdamped Langevin equations. Practically, the updated positions at time step $i+1$ are computed from the positions at time step $i$ according to

$$\mathbf{r}_k^{(i+1)} = \mathbf{r}_k^{(i)} + \frac{\Delta t}{6\pi\eta a}\mathbf{F}_k^{(i)} + \sqrt{\frac{2k_\mathrm{B}T\Delta t}{6\pi\eta a}}\mathbf{X}_k^{(i)}, \quad (4)$$

where $\Delta t = 10$ ps is the integration time step, $\eta = 0.00089$ Pa×s the viscosity of water, $\mathbf{F}_k^{(i)}$ the force arising from potential energy $V_\mathrm{DNA} + V_\mathrm{wall}$ acting on bead $k$, and $\mathbf{X}_k^{(i)}$ a vector of random numbers extracted from a Gaussian distribution of mean 0 and variance 1. Forces $\mathbf{F}_k^{(i)}$ are quite standard and easy to calculate, except for the torsion contribution, for which Eq. A23, A26 and A27 of [19] are used.

Updated body-fixed orthogonal frames $(\mathbf{f}_k^{(i+1)}, \mathbf{v}_k^{(i+1)}, \mathbf{u}_k^{(i+1)})$ are computed along slightly different lines compared to [19], although the underlying concept is the same and respective trajectories differ only marginally. The aim is to allow for a more straightforward description of the polymerase (see below). The procedure is described here step by step. $\beta_k^{(i)}$ and $\alpha_k^{(i)} + \gamma_k^{(i)}$ are obtained from

$$\cos(\beta_k^{(i)}) = \mathbf{u}_k^{(i)} \cdot \mathbf{u}_{k+1}^{(i)}, \quad (5)$$

$$\cos(\alpha_k^{(i)} + \gamma_k^{(i)}) = (\mathbf{f}_k^{(i)} \cdot \mathbf{f}_{k+1}^{(i)} + \mathbf{v}_k^{(i)} \cdot \mathbf{v}_{k+1}^{(i)})/[1 + \cos(\beta_k^{(i)})], \quad (6)$$

$$\sin(\alpha_k^{(i)} + \gamma_k^{(i)}) = (\mathbf{v}_k^{(i)} \cdot \mathbf{f}_{k+1}^{(i)} - \mathbf{f}_k^{(i)} \cdot \mathbf{v}_{k+1}^{(i)})/[1 + \cos(\beta_k^{(i)})]. \quad (7)$$

According to its definition, the updated unit vector $\mathbf{u}_k^{(i+1)}$ is just

$$\mathbf{u}_k^{(i+1)} = (\mathbf{r}_{k+1}^{(i+1)} - \mathbf{r}_k^{(i+1)})/|\mathbf{r}_{k+1}^{(i+1)} - \mathbf{r}_k^{(i+1)}|. \quad (8)$$

Rotation $\mathrm{R}[\mathbf{N}_k^{(i)}, \delta\psi_k^{(i)}]$, where

$$\mathbf{N}_k^{(i)} = (\mathbf{u}_k^{(i)} \times \mathbf{u}_k^{(i+1)})/|\mathbf{u}_k^{(i)} \times \mathbf{u}_k^{(i+1)}|, \quad (9)$$

$$\delta\psi_k^{(i)} = \arcsin(|\mathbf{u}_k^{(i)} \times \mathbf{u}_k^{(i+1)}|), \quad (10)$$

transforms $\mathbf{u}_k^{(i)}$ into $\mathbf{u}_k^{(i+1)}$. The image of $\mathbf{f}_k^{(i)}$ by this rotation,

$$\mathbf{g}_k^{(i)} = \mathrm{R}[\mathbf{N}_k^{(i)}, \delta\psi_k^{(i)}](\mathbf{f}_k^{(i)}), \quad (11)$$

is computed according to Rodrigues' formula

$$\mathbf{g}_k^{(i)} = \cos(\delta\psi_k^{(i)})\,\mathbf{f}_k^{(i)} + [1 - \cos(\delta\psi_k^{(i)})][\mathbf{N}_k^{(i)} \cdot \mathbf{f}_k^{(i)}]\mathbf{N}_k + \sin(\delta\psi_k^{(i)})(\mathbf{N}_k^{(i)} \times \mathbf{f}_k^{(i)}). \quad (12)$$

$\mathbf{g}_k^{(i)}$ takes into account the infinitesimal rotation of $\mathbf{f}_k^{(i)}$ due to the local reorientation of the DNA chain between steps $i$ and $i+1$. The modulus of the total torque exerted on bead $k$ is next obtained according to

$$T_k^{(i)} = \tau(\alpha_k^{(i)} + \gamma_k^{(i)} - \alpha_{k-1}^{(i)} - \gamma_{k-1}^{(i)}), \quad (13)$$

where $\tau$ is the torsion force constant introduced in Eq. 1. $\mathbf{f}_k^{(i+1)}$ is the image of $\mathbf{g}_k^{(i)}$ by a second rotation $\mathrm{R}[\mathbf{u}_k^{(i+1)}, \delta\phi_k^{(i)}]$, where

$$\delta\phi_k^{(i)} = \frac{\Delta t}{4\pi\eta a^2 l_0}T_k^{(i)} + \sqrt{\frac{2k_\mathrm{B}T\Delta t}{4\pi\eta a^2 l_0}}Y_k^{(i)}, \quad (14)$$

and $Y_k^{(i)}$ is a random number extracted from a Gaussian distribution of mean 0 and variance 1. One gets

$$\mathbf{f}_k^{(i+1)} = \mathrm{R}[\mathbf{u}_k^{(i+1)}, \delta\phi_k^{(i)}](\mathbf{g}_k^{(i)}) = \cos(\delta\phi_k^{(i)})\,\mathbf{g}_k^{(i)} + \sin(\delta\phi_k^{(i)})\,(\mathbf{u}_k^{(i+1)} \times \mathbf{g}_k^{(i)}). \quad (15)$$

Note that the small correlations between translational and rotational noise, which arise from the fact that DNA beads are treated as cylinders in Eq. 14 and from the proximity of other DNA beads, are neglected in Eqs. 4 and 14. The third component of the updated body-fixed orthogonal frame is finally computed according to

$$\mathbf{v}_k^{(i+1)} = \mathbf{u}_k^{(i+1)} \times \mathbf{f}_k^{(i+1)}. \quad (16)$$

A transcribing polymerase is modeled as a transient modification of the evolution equations of the DNA bead on which it sits at a given time step $i$. More precisely, if the polymerase sits on bead $\Gamma$, then this bead behaves as a large sphere of hydrodynamic radius $A = 6$ nm, which is the approximate dimension of a single RNAP enzyme [26,27]. $\mathbf{r}_\Gamma^{(i+1)}$ is consequently computed according to

$$\mathbf{r}_\Gamma^{(i+1)} = \mathbf{r}_\Gamma^{(i)} + \frac{\Delta t}{6\pi\eta A}\mathbf{F}_\Gamma^{(i)} + \sqrt{\frac{2k_\mathrm{B}T\Delta t}{6\pi\eta A}}\mathbf{X}_\Gamma^{(i)}, \quad (17)$$

instead of Eq. 4. Moreover, $\delta\phi_{\Gamma}^{(i)}$ is expressed in the form

$$\delta\phi_{\Gamma}^{(i)} = \frac{\Delta t}{8\pi\eta A^3} T_{\Gamma}^{(i)} + \sqrt{\frac{2k_{\mathrm{B}}T\Delta t}{8\pi\eta A^3}}\, Y_{\Gamma}^{(i)} - \Omega\Delta t, \tag{18}$$

instead of Eq. 14. In addition to the modification of the rotational diffusion constant in the first two terms, the right-hand side of Eq. 18 displays a third term, $-\Omega\Delta t$, which represents the energy-consuming action of the motor, in the spirit of [28-30]. For positive values of $\Omega$, the additional term increases the local density of twist for beads $k > \Gamma$ ("downstream") and decreases it for beads $k < \Gamma$ ("upstream"). Translocation of the motor is modeled by keeping track of the total rotation angle $\Phi$ of the body-fixed frame at bead $\Gamma + 10$ since time step $I$ when the last translocation event occurred :

$$\Phi = \sum_{i>I} \delta\phi_{\Gamma+10}^{(i)}. \tag{19}$$

Since each bead represents 7.5 bp and the number of base pairs per DNA helix turn is 10.5, the polymerase translocates forward (i.e., $\Gamma$ is incremented by $+1$) when $\Phi$ reaches $-15\pi/10.5$ or backward (i.e., $\Gamma$ is incremented by $-1$) when $\Phi$ reaches $+15\pi/10.5$. Rotation angle since last translocation occured is computed at bead $\Gamma + 10$ instead of $\Gamma$, because the torsion angle between beads $k$ and $k+1$ is discontinuous at $\Gamma$ (Figs S2 of [17]) and one wants to avoid eventual fluctuations associated with this discontinuity. In the present work, as in [17] and [18], $\Omega$ was taken as a free parameter, which was varied to investigate different regimes.

At any time step, the excess of twist, $\Delta T\mathrm{w}$, the writhe, $W\mathrm{r}$, the linking number difference, $\Delta L\mathrm{k}$, and the superhelical density, $\sigma$, are computed according to

$$\Delta T\mathrm{w} = \frac{1}{2\pi}\sum_k (\alpha_k + \gamma_k), \tag{20}$$

$$W\mathrm{r} = \frac{1}{4\pi}\sum_k \sum_{m \neq k} \frac{[(\mathbf{r}_{m+1} - \mathbf{r}_m) \times (\mathbf{r}_{k+1} - \mathbf{r}_k)] \cdot (\mathbf{r}_m - \mathbf{r}_k)}{|\mathbf{r}_m - \mathbf{r}_k|^3}, \tag{21}$$

$$\Delta L\mathrm{k} = \Delta T\mathrm{w} + W\mathrm{r}, \tag{22}$$

$$\sigma = \frac{\Delta L\mathrm{k}}{L\mathrm{k}_0}, \tag{23}$$

where

$$L\mathrm{k}_0 = \frac{7.5n}{10.5}. \tag{24}$$

$\Delta L\mathrm{k}$ and $\sigma$ are topological constants and are expected not to vary as long as the circular chain is not opened. This point is used as an internal check of the fact that electrostatic repulsion in Eq. 1 is sufficiently strong for different segments of the DNA chain not to cross. Any crossing would result in a sudden jump of $\Delta L\mathrm{k}$ by $\pm 2$. By continuously monitoring $\Delta L\mathrm{k}$, we ascertain that this never happens.

Of interest are also the local excess of twist at bead $k$,

$$\Delta T\mathrm{w}(k) = \frac{1}{2\pi}(\alpha_k + \gamma_k), \tag{25}$$

as well as the local writhe at bead $k$,

$$W\mathrm{r}(k) = \frac{1}{4\pi}\sum_{m=k-L}^{m=k-1} \sum_{j=k}^{j=k+L-1} \frac{[(\mathbf{r}_{m+1} - \mathbf{r}_m) \times (\mathbf{r}_{j+1} - \mathbf{r}_j)] \cdot (\mathbf{r}_m - \mathbf{r}_j)}{|\mathbf{r}_m - \mathbf{r}_j|^3}, \tag{26}$$

which provides a measure of the self-crossing of segment $[k-L, k+L]$ [17,18,31-33]. $L$ was set to 1.5 times the DNA persistence length $L_{\mathrm{p}} \approx 50$ nm [20], that is $L = 30$ beads.

The experimental conditions of Figure 1, where a single DNA track is heavily transcribed, were modeled by assuming that a single polymerase translocates repeatedly along a given DNA track. When the polymerase reaches the end of the track, it is re-positioned immediately at the beginning of the track and a new cycle begins. Unless otherwise stated, the results discussed in this paper were obtained with a transcription track composed of 150 beads (1125 bp). Several sets of simulations were performed with a longer track composed of 450 beads (3375 bp), in order to check the impact of the length of the transcription track on the contact map.

Contact maps were computed by checking every $10^5$ time steps (1 µs time intervals) whether the centers of beads $k$ and $m$ are separated by less than 10 nm and computing the corresponding contact probability over 100 successive time intervals (100 µs time windows). Density maps showing the logarithm to the base 10 of the contact probability as a function of $k$ and $m$ were subsequently plotted for each time window. In these contact maps, plectonemes appear as segments parallel to the secondary diagonal of the map. Note that the experimental capture radius is usually in the range 100-200 nm [34], rather than 10 nm, with the consequence that the resolution of the computed contact maps discussed in the present work is significantly better than that of experimental ones. For the sake of clarity, we preferred not to degrade the resolution of computed contact maps down to that of experimental ones.

The combination of parameters (track length, DNA superhelicity and rate of twist injection), which has been explored, is summarized in Table 1. Let us mention that the characteristic contact pattern discussed in the present work results from the injection of a large quantity of twist in the DNA chain, which generates a large and uniform torque throughout the DNA chain. As a consequence, noise plays only a minor role and simulations are reproducible, except for some limited jitter concerning the precise time at which a given phenomenon occurs. All simulations were run for a sufficiently long time for this jitter not to impact the discussions below.

## 3. Results

### *3.1. Threshold value of $\Omega/n$*

The value of the rate of twist injection $\Omega$ introduced in Eq. 18 is not known *a priori*. Reasonable values thereof can be estimated by comparing the results of BD simulations with those of experiments. From this point of view, Figure 3(B) of [16] is of particular interest. This figure shows that plectonemes form on both sides of a polymerase that transcribes a 31 kbp torsionally relaxed ($\sigma = 0$) DNA molecule, whose ends are tightly bound to a planar surface. This indicates that the torque induced by transcription exceeds the buckling threshold. Since in the CG model the average torque exerted on each bead is proportional to $\Omega/n$ [17,18], this experimental result may be used to set a lower bound to $\Omega/n$. Indeed, as can be checked in the contact maps shown in the left column of Figure 2, simulations performed with $\sigma = 0$ and $\Omega/n \leq 500$ rad×s$^{-1}$×bead$^{-1}$ display only small and short-lived structures, whereas those performed with $\Omega/n \geq 1000$ rad×s$^{-1}$×bead$^{-1}$ display instead long and stable plectonemes, as well as additional features that will be discussed below. This comparison between experiments and simulations therefore suggests that the action of a transcribing polymerase is adequately modeled when using values of $\Omega/n$ larger than a threshold value comprised between 500 and 1000 rad×s$^{-1}$×bead$^{-1}$ (note that this threshold is a simulation-derived quantity rather than an independent torque measurement, which would also depend on tension).

Strikingly, new features also appear at $\Omega/n = 1000$ rad×s$^{-1}$×bead$^{-1}$ in contact maps computed for $\sigma = -6.3$ %, as can be checked in the right column of Figure 2. For this value of $\sigma$, which is quite usual in bacteria [35,36], DNA molecules form branched plectonemes [21,37]. Examination of Figure 2 indicates that, up to $\Omega/n = 500$ rad×s$^{-1}$×bead$^{-1}$, the polymerase sits most of the time at, or close to, the apex of a long plectoneme, a preferred localization already reported in previous independent studies [38,39]. In contrast, when $\Omega/n \geq 1000$ rad×s$^{-1}$×bead$^{-1}$, the polymerase sits close to the foot of a long plectoneme, which is itself part of a larger pattern that closely resembles the arched stripe of Figure 1.

In conclusion, these results indicate that the arched stripe pattern is observed in computed contact maps of underwound DNA chains for values of $\Omega/n$ that realistically model transcribing polymerases.

### *3.2. The arched stripe pattern in BD simulations*

Figure 3 shows the contact map computed from a simulation with $\sigma = -6.3$ % and $\Omega/n = 2000$ rad×s$^{-1}$×bead$^{-1}$ at $t = 20$ ms, that is, after the polymerase has translocated 9 times along the 150 beads track delimited by the gray lines. Comparison of Figures 1 and 3 indicates that the arched stripe pattern labeled (1) in Figure 1 actually decomposes into a contact line due to a plectoneme, which is labeled $(p)$ in Figure 3, plus a contact line labeled $(b)$, which is tilted with respect to $(p)$. An additional curved contact line, labeled $(a)$ in Figure 3, connects the two ends of $(p)$. The $(a)$ contact line looks like a "*counter-arch*", which opposes the arch formed by $(p)$ and $(b)$. Closer inspection of Figure 1 reveals that the counter-arch is also present in the experimental contact map (it is labeled (3) in Figure 1), although it was not specifically discussed in [15].

In Figure 3, the arched stripe extends over more than half the length of the DNA chain. Complete simulations indicate that its size increases almost linearly with time and that it extends over all the DNA chain (43200 bp) at about $t = 43$ ms. If simulations had been performed with a longer DNA chain, then the arched stripe would probably have grown beyond 40 kbp. This is consistent with the experimental contact map in Figure 1, where the arched stripe pattern extends over about 60 kbp. Straight extrapolation suggests that the 60 kbp size would be attained in about 60 ms real time, but computer time would become prohibitively long for longer DNA chains with $n = 12000$ or 18000 beads. It is anyway worth noting that the characteristic size of the arched stripe is more than one order of magnitude larger than the characteristic size of a gene (about 1000 bp) or an operon (a few genes), and almost one order of magnitude larger than the average size of supercoiled branches (on average 10 kbp in *E. coli* chromosome [40] and about 5-6 kbp for the shorter DNA chains investigated here). In fact, the only length scale with which the size of the arched stripe correlates well is the width of the supercoiled domain generated by transcription, which has been shown to span up to 25 kbp on both sides of a transcribing RNAP [4]. We will come back to this point in Section 4.1.

### *3.3. Entangled DNA geometry*

The peculiar geometry of the DNA coil resulting in the arched stripe and counter-arch patterns in contact maps was unveiled by plotting the conformation of the DNA coil (Figure 4), as well as the local excess of twist and the local writhe (Figure 5). Note that Figures 3, 4 and 5 were derived from the same trajectory. In Figure 4, DNA segments involved in contacts generated by transcription are shown as colored tubes, while the rest of the DNA chain is shown as a gray transparent tube. The same color code is used in Figure 5 to identify the various segments more easily.

In an equilibrated DNA chain with superhelical density $\sigma = -6.3$ %, the excess of twist is homogeneously distributed along the chain, with a uniform local excess of twist $\Delta T\mathrm{w}(k) \approx -1.3$ %. Moreover, the chain forms a branched plectoneme, where all plectonemic segments are negatively supercoiled. This is no longer the case when a polymerase transcribes the DNA chain and continuously injects waves of positive supercoiling downstream of its position and negative supercoiling upstream of it (modeled by the $-\Omega\Delta t$ term in Eq. 18). The continuous injection of twist generates at steady state a uniform torque along the DNA chain, or equivalently a constant gradient of torsion angles between successive beads (Eq. 13) and a constant gradient of local twist (Eq. 25). This gradient is compensated by a discontinuity at the position of the polymerase, where twist is injected (Fig. S2 of [17])). As can be checked in the bottom panel of Figure 5, $\Delta T\mathrm{w}(k)$ decreases indeed linearly with $k$ and is discontinuous at $k = \Gamma$, the position of the polymerase. The local excess of twist reaches $\Delta T\mathrm{w}(k) \approx -4.0$ % immediately upstream of the polymerase, but it is positive and as large as $\Delta T\mathrm{w}(k) \approx 2.0$ % immediately downstream of it, the mean of these two limits being close to the uniform value in the absence of polymerase. As a consequence, a region with $\Delta T\mathrm{w}(k) \approx 0$ is observed ap-proximately opposite to the polymerase position. Remembering that a narrow peak in the plot of $Wr(k)$ corresponds to the apex of a plectoneme [17,18,31-33], the top panel of Figure 5 indicates that the local writhe adapts to the variations of the local excess of twist. Indeed, negatively supercoiled plectonemes accumulate upstream of the polymerase and positively supercoiled plectonemes downstream of it, while the region of the chain opposite to the polymerase is no longer plectonemic. In Figure 4, the polymerase consequently sits at the common foot of a long negatively supercoiled plectoneme labeled $(P^-)$ (shown in brown in panels A and B of Figure 4) and a much shorter positively supercoiled plectoneme labeled $(P^+)$ (shown in red in panels A and D of Figure 4). Contacts within $(P^-)$, which is located in the range $[\Gamma - 1120, \Gamma]$, result in the $(p)$ contact line in Figure 3. Moreover, the $(a)$ contact line is due to the fact that $(P^-)$ toroidally winds around a segment labeled $(A^+)$, which is located downstream of the polymerase, in the range $[\Gamma + 180, \Gamma + 310]$. $(A^+)$ is shown in green in panels A and B of Figure 4. The negative toroidal winding of $(P^-)$ around $(A^+)$ also appears as an irregular plateau at $Wr(k) \approx -0.2$ in the top panel of Figure 5. The central part of the arched stripe and the counter-arch therefore reflect the fact that the polymerase sits at one foot of a long negatively supercoiled plectoneme, which winds around a short DNA segment located downstream of the polymerase.

The edges of the arched stripe reflect yet another level of complexity of the geometry of the DNA coil. Indeed, the $(b)$ contact line in Figure 3 is due to the long DNA segment labeled $(B^+)$ and highlighted in yellow in panels A and C of Figure 4, which toroidally winds around the segment labeled $(B^-)$ and shown in black. $(B^+)$ is located downstream of the polymerase, in the range $[\Gamma + 310, \Gamma + 1670]$. In the absence of transcription, it would be negatively supercoiled. Because the polymerase continuously injects positive twist downstream of its position, the negatively supercoiled plectonemes are however destabilized and replaced by short positively supercoiled ones, as can be checked in the top panel of Figure 5. In contrast, the $(B^-)$ DNA segment is located upstream of both the polymerase and the $(P^-)$ plectoneme, in the range $[\Gamma - 1530, \Gamma - 1180]$. Positive toroidal winding of $(B^+)$ around $(B^-)$ also appears as an irregular plateau at $Wr(k) \approx 0.25$ in the top panel of Figure 5. Since $(B^+)$ is about 4 times longer than $(B^-)$, the $(b)$ contact line is tilted with respect to the $(p)$ one, with the union of $(p)$ and $(b)$ forming the arched stripe.

In conclusion, the arched stripe and counter-arch patterns observed in Hi-C contact maps reflect the rather entangled geometry of the DNA molecule around highly transcribed DNA loci, with DNA tracts located upstream of the polymerase winding around tracts located downstream, and *vice-versa*.

### *3.4. Dynamical aspects*

As already mentioned, an equilibrated DNA chain with superhelical density $\sigma = -6.3$ % forms a branched plectoneme. The position and the length of plectonemic branches do not remain constant. Instead, plectonemes diffuse along the DNA molecule and undergo growth/shrinkage events, during which a plectoneme at a certain position shrinks (and eventually disappears), while a plectoneme at another, possibly distant position simultaneously grows [16,18,41,42]. When transcription starts, the polymerase is consequently located at a random position in a plectonemic branch. This is no longer the case during transcription. Indeed, as can be checked in the right column of Figure 2, the polymerase systematically sits close to the apex of a long plectoneme if $\Omega/n \leq 500$ rad×s$^{-1}$×bead$^{-1}$, whereas it localizes close to the foot of a plectoneme if $\Omega/n \geq 1000$ rad×s$^{-1}$×bead$^{-1}$. Transition from the initial to the final position occurs through a series of coordinated diffusion and growth/shrinkage events [18], which absorb little energy. For $\Omega/n \geq 1000$ rad×s$^{-1}$×bead$^{-1}$, the net result is that the $(P^-)$ plectoneme grows progres-sively at the upstream side of the polymerase, which is reflected in the corresponding growth of the $(p)$, $(a)$ and $(b)$ lines in contact maps. This is clearly seen in Figure 6, which shows superposed contact maps computed at $t = 0$, 10, 20 and 30 ms from a simulation with $\sigma = -6.3$ % and $\Omega/n = 2000$ rad×s$^{-1}$×bead$^{-1}$. Growth is almost linear with time, as can be checked in Figure 7, which shows the time evolution of $k_{\text{apex}}$, the position of the apex of $(P^-)$. The red solid line corresponds to the simulation with $\sigma = -6.3$ % and $\Omega/n = 2000$ rad×s$^{-1}$×bead$^{-1}$. It is seen that $k_{\text{apex}}$ decreases regularly from about 2850 down to about 2000 in 43 ms. At this latter time, $(P^-)$ is about 1700 beads long and the toroidal winding of $(B^+)$ around $(B^-)$ accounts for the rest of the DNA chain.

A finer understanding of how growth proceeds is gained by looking at Figure 8, which shows contact maps computed at $t = 7.2$, 7.7, and 9.5 ms from the simulation with $\sigma = -6.3$ % and $\Omega/n = 2000$ rad×s$^{-1}$×bead$^{-1}$. At $t = 7.2$ ms (top left panel of Figure 8), the polymerase reaches the end of the transcription track. It sits at one foot of $(P^-)$, which consequently coincides with the end of the track. Immediately after reaching the end of the track, the polymerase is repositioned at its beginning and starts again moving forward. At $t = 7.7$ ms (top right panel of Figure 8), the polymerase has moved forward by 50 beads. Most importantly, the DNA segment located between the polymerase and the end of the transcription track (thick orange arrow) has been destabilized and is no longer plectonemic. Indeed, the corresponding piece of contact line is no longer parallel to the secondary diagonal of the plot, like the $(p)$ contact line, but is instead tilted like the $(b)$ one. Stated in other words, this segment henceforth belongs to $(B^-)$ instead of $(P^-)$. As the polymerase moves forward, $(P^-)$ goes on growing, thanks to coordinated diffusion and growth/shrinkage events [18]. The net result is that, at $t = 9.5$ ms, as the polymerase again reaches the end of the transcription track (bottom left panel of Figure 8), $(P^-)$ extends again to the end of the track but is longer than at $t = 7.2$ ms. In addition, the piece of $(P^-)$, which at $t = 7.2$ ms extended from the beginning to the end of the transcription track, has been transferred to $(B^-)$ and henceforth contributes to the edges of the arched stripe instead of its center. The evolution of the contact map after the polymerase has completed one additional translocation round is best seen in the bottom right panel of Figure 8, where the contact maps at $t = 7.2$ and 9.5 ms are superposed.

### *3.5. Bundled domain*

The discussion above focused on the arched stripe and counter-arch patterns, which are respectively labeled (1) and (3) in Figure 1. The authors of the experimental study additionally report a bundled domain, which is labeled (2) in Figure 1 [15]. This pattern may be longer than 80 kbp, but is "only" 20 kbp wide, that is significantly less than the arched stripe itself. It is located downstream of the polymerase, in a region which is negatively supercoiled in the absence of transcription but positively supercoiled when transcription is active (bottom panel of Figure 5). According to simulations, this region contains several positively supercoiled plectonemes, which are significantly shorter than $(P^-)$ and responsible for the contact lines labeled $(p^*)$ in Figure 3. The positively supercoiled plectoneme located closest to the polymerase, in the range $[\Gamma, \Gamma + 180]$, is labeled $(P^+)$ and shown in red in panel D of Figure 4. It toroidally winds around a DNA segment located upstream of the polymerase, in the range $[\Gamma - 1180, \Gamma - 1120]$, which is labeled $(C^-)$ and shown in blue in panel D of Figure 4. The toroidal winding of $(P^+)$ around $(C^-)$ is responsible for the short, rounded contact line labeled $(c)$ in Figure 3. Three other short positively supercoiled plectonemes belong to $(B^+)$, which winds around $(B^-)$ (panel C of Figure 4).

Most importantly, as transcription proceeds, short positively plectonemes are regularly born at the downstream side of the polymerase and grow for a short time, before they separate and travel away from it. This is clearly seen in Figure 9, which shows the time evolution of the local writhe $Wr(k)$ for all DNA beads $k$. In this figure, red (respectively, blue) lines denote the position of the apex of positively (respectively, negatively) supercoiled plectonemes. According to Figure 9, in about 40 ms, seven positively supercoiled plectonemes are born at the downstream side of the polymerase and travel away at a speed of about 25 beads×ms$^{-1}$. Noteworthy, plectonemes preserve their shape and length while travelling over more than thousand beads (10 kbp).

The bundled domain reported in [15] and labeled (2) in Figure 1 is most probably the trace in experimental contact maps of these travelling, short, and positively supercoiled plectonemes generated by polymerase translocation.

## 4. Discussion

### *4.1. The arched stripe pattern in wild type systems ?*

The contact pattern in Figure 1 was obtained under quite peculiar conditions. Indeed, a T7 promoter was inserted in the genome of *Escherichia coli* cells and T7 RNAP enzymes were introduced in cells where endogenous RNA synthesis was blocked by rifampicin [15]. One may consequently wonder whether the arched stripe pattern is specific to this peculiar system or whether there is some hope to observe it in a wild type system, eventually in vivo. Since this work suggests that the arched stripe pattern is a threshold effect driven by the rate of twist injection, this question actually amounts to asking whether some bacterial genes are transcribed at a larger rate than the typical transcription rate of the T7 RNAP enzyme. In fact, this is indeed the case. For example, under moderate to rich conditions, ribosomal RNA genes (rrn) in *E. coli* are transcribed about two orders of magnitude more frequently than a typical T7-driven gene (about 50 initiations per minute per gene [43] against 0.5 initiations per minute per gene [44]). What is most interesting is that the supercoiled domain generated by the transcription of ribosomal genes has been studied in detail [4] and displays striking similarities with the present work. In particular, it spans about 25 kbp on both sides of the transcribing RNAP. Moreover, its "amplitude", that is the gap of psoralen binding on both sides of the polymerase, which is directly proportional to the gap of twist, increases linearly with gene expression for highly expressed genes [4]. Particularly satisfying is the comparison of Figure 4B of [4] with the bottom plot of Figure 5. Finally, the seven ribosomal gene operons represent together as much as 70% of transcriptional activity under rapid growth conditions [45], which implies that the deformations imposed to the DNA molecule by their transcription are not likely to be obscured by the activity of neighboring genes. Taken together, these arguments suggest that the arched stripe contact pattern may eventually be observed around ribosomal genes under moderate to rich conditions.

### *4.2. Toroidal supercoiling is an out-of-equilibrium feature*

The present work furthermore suggests that toroidal supercoiling is, in the context of transcription, an out-of-equilibrium feature that requires continuous injection of twist, whereas plectonemic supercoiling is essentially an equilibrium feature. Indeed, the circular DNA chain forms at equilibrium a branched plectoneme and polymerase translocation alters only to a limited extent the total amount of plectonemic writhe. In contrast, toroidal supercoiling builds up only when the polymerase translocates along the DNA chain and disappears rapidly when translocation ceases. This point can be checked in Figure 10, which shows two superposed contact maps. The red one is the contact map already shown in Figure 3, which is computed at $t = 20$ ms from a simulation with $\sigma = -6.3$ % and $\Omega/n = 2000$ rad×s$^{-1}$×bead$^{-1}$. The green map was obtained from the red one by letting the system evolve for an additional 6 ms in the absence of polymerase translocation, that is with $\Omega/n = 0$ rad×s$^{-1}$×bead$^{-1}$. It is seen that the $(p)$ contact line, and consequently also the $(P^{-})$ plectoneme, evolve very little during these additional 6 ms.

In contrast, the $(a)$ contact line (the counter-arch) has almost disappeared in the green map, which means that $(P^-)$ no longer winds around a segment of DNA located downstream of the polymerase. Still more impressive is the fact that the several fragments which compose the $(b)$ contact line have re-oriented parallel to the secondary diagonal of the map, which means that the toroidal winding of $(B^+)$ around $(B^-)$ has transformed into a more usual branched plectoneme.

*4.3. Influence of $\sigma$, $\Omega$ and track length*

Discussion in Section 3 and illustrations in Figures 3 to 10 are based on simulations performed with $\sigma = -6.3$ %, $\Omega/n = 2000$ rad×s$^{-1}$×bead$^{-1}$ and a 150 beads transcription track. How the contact pattern around highly transcribed sites evolves for different values of these parameters is analyzed in some detail in Appendix A. Most relevant conclusions are summarized in the present subsection.

Superhelical density $\sigma$ plays a crucial role on both the location and stability of the contact pattern. The reason, why for $\sigma = -6.3$ % the long $(P^-)$ plectoneme forms systematically on the upstream side of the polymerase, is that the waves of supercoiling generated by the translocating polymerase reinforce plectonemic branches located upstream of its position but destabilize those located downstream. For $\sigma = 0$ %, the DNA chain is instead torsionally relaxed in the absence of polymerase activity and there is equal probability that polymerase translocation generates positively supercoiled plectonemes downstream of its position or negatively supercoiled ones upstream of it. As a consequence, the arched stripe may form both upstream and downstream of the polymerase, pointing towards opposite directions. For example, a positively supercoiled plectoneme and the associated counter-arch are clearly seen downstream of the polymerase in the bottom left panel of Figure 2. Simulations indicate that the threshold where arched stripes may form on both sides of the polymerase lies somewhere between $\sigma = -0.8$ % and $\sigma = -1.6$ % (fifth column of Table 1). Less negative values of $\sigma$ are also associated with instability, in the sense that $(P^-)$ prematurely detaches and travels away from the upstream side of the polymerase instead of growing almost nearly till the contact pattern fills the whole DNA chain, as is the case for more negative values of $\sigma$. The threshold for stability lies somewhere between $\sigma = -1.6$ % and $\sigma = -2.4$ % (sixth column of Table 1).

As discussed in Section 3.1, the specific contact pattern around highly transcribed genes is observed only above a threshold value which lies somewhere between $\Omega/n = 500$ rad×s$^{-1}$×bead$^{-1}$ and $\Omega/n = 1000$ rad×s$^{-1}$×bead$^{-1}$. Simulations furthermore indicate that, above the threshold, the growth rate of the arched stripe and counter-arch patterns is nearly proportional to $\Omega$ (seventh column of Table 1). This may be checked in Figure 7, where the slope of the blue dashed line obtained with $\Omega/n = 1000$ rad×s$^{-1}$×bead$^{-1}$ is approximately one half of the slope of the brown and red lines obtained with $\Omega/n = 2000$ rad×s$^{-1}$×bead$^{-1}$.

Let us finally mention that longer transcription tracks also result in some instability of the contact pattern, because the polymerase moves too fast and over too long a distance for $(P^-)$ to follow its motion [18]. For the 450 beads track, $(P^-)$ detaches from the upstream side of the polymerase somewhere near the middle of the track and connects again to it when the polymerase starts a new translocation from the beginning of the track. This intermittent process is less efficient than the continuous process observed for the 150 beads track and results in a smaller apparent growth rate of the contact pattern.

*4.4. "Jets" and "fountains" arising from loop extrusion by cohesin*

It is worth noting that the contact pattern in Figure 1 is quite similar to those displayed, for example, in Figure 7 of [46] and Figures 3 and 6 of [47], but that their origin is, however, totally different. Indeed, the contact patterns in [46] and [47], which have been called “jets” or “fountains”, are thought to arise from focal loading and subsequent loop extrusion by cohesin. They appear as contact signals that emanate from a relatively narrow genomic locus and broaden with increasing genomic distance, consistent with cohesin complexes loaded at a preferential site and extruding chromatin bidirectionally. In quiescent mammalian lymphocytes, such “jets” can extend over megabase scales [46]. Most interestingly, binding of CFTC (a protein that acts as a stop sign and blocks extrusion in a specific direction) affects jet angle and consequently the shape of the arrow head [46]. In *C. elegans* and zebrafish, these structure have been called “fountains”, but depend also from cohesin loading and processive extrusion [47]. More generally, the extent of these structures reflects the balance between cohesin loading, extrusion processivity, removal of the cohesin complex by the WAPL regulatory protein, and CTCF barriers [48].

## 5. Conclusions

The CG model has proved rather helpful to decipher the unusual features of high-resolution Hi-C contact maps around highly expressed loci. Part of this success is due to the fact that the level of graining is coarse enough to investigate DNA molecules that are tens of kbp long over sufficiently long times, while still fine enough to adequately describe DNA rigidity and writhing. As a result, a one-to-one correspondence could be established between the details of experimental contact maps and the underlying conformation of the DNA coil, thereby highlighting the extraordinarily entangled geometry of the DNA molecule. Most interestingly, the authors of the contact map in Figure 1 also recorded high-resolution contact maps around two neighboring loci being simultaneously transcribed (Figure 3 of [15]). From the theoretical point of view, it is known that the waves of supercoiling generated by the expression of a gene impact the expression of neighboring ones [49-53]. A natural extension of the present work would therefore be to use the CG model to understand whether the interactions between genes being simultaneously transcribed explain some of the features of the complex contact maps displayed in Figure 3 of [15]. Another interesting development would be to investigate how gene transcription impacts the localization and, eventually, detachment of neighboring nucleosomes in eukaryotic cells [54,55]. Anyway, it is anticipated that the kind of modeling approach used in the present work will increasingly help Chromosome Conformation Capture experiments decipher the internal organization of bacterial nucleoids, especially if observation of the time evolution of contact maps becomes feasible shortly [56].

**Funding:** This research received no external funding.

**Conflicts of Interest:** The author declares no conflicts of interest.

## Appendix A.

*Appendix A.1. Influence of superhelical density* $\sigma$

For $\sigma = -6.3$ %, the contact pattern is very asymmetric, with a long negatively supercoiled plectoneme growing at the upstream side of the polymerase and short positively supercoiled ones detaching regularly from its downstream side and traveling away. This asymmetry is due to the fact that the wave of negative (respectively, positive) supercoiling generated by the polymerase upstream (respectively, downstream) of its position reinforces (respectively, destabilizes) the negatively supercoiled plectonemes located on this side of the polymerase. No asymmetry is instead expected for torsionally relaxed DNA ($\sigma = 0$ %), especially as overwinding and underwinding the DNA chain are fully equivalent in the CG model. Simulations indeed confirm that for $\sigma = 0$ % the probability that a long positively supercoiled plectoneme grows at the downstream side of the polymerase is equal to the probability that a long negatively supercoiled one grows at its upstream side. This is illustrated in the bottom left panel of Figure 2, where a positively supercoiled plectoneme is clearly seen downstream of the polymerase. The counter-arch is also present, which means that the plectoneme winds around a segment of DNA located upstream of the polymerase. Additional simulations performed with $\sigma = -0.8\ \%$ , $-1.6\ \%$ , $-2.4\ \%$ and $-3.2\ \%$ reveal that the threshold where arched stripes may form on both sides of the polymerase lies somewhere between $\sigma = -0.8\ \%$ and $\sigma = -1.6\ \%$ (fifth column of Table 1). Moreover, less negative values of $\sigma$ are associated with instability, in the sense that $(P^-)$ prematurely detaches and travels away from the upstream side of the polymerase instead of steadily growing (sixth column of Table 1). The threshold for stability lies somewhere between $\sigma = -1.6\ \%$ and $\sigma = -2.4\ \%$ (sixth column of Table 1). Moreover, the growth rate of $(P^-)$ is similar for the three values of $\sigma$, as can be checked in the seventh column of Table 1, as well as in Figure 7, where the brown dotted line shows the time evolution of $k_{\mathrm{apex}}$ for a simulation with $\sigma = -3.2\ \%$. In contrast, the growth rate of the $(b)$ contact line is significantly larger for more negative values of $\sigma$, as can be checked in the last column of Table 1, as well as in Figure A1, which shows superposed contact maps computed at $t = 25$ ms from simulations with $\Omega/n = 2000$ rad×s$^{-1}$×bead$^{-1}$ and $\sigma = -6.3\ \%$ (red map) or $\sigma = -2.4\ \%$ (green map). This indicates that a more negative superhelical density favors the winding of DNA segments located downstream of the polymerase around segments located upstream.

*Appendix A.2. Influence of the rate of twist injection $\Omega$*

As discussed in Section 3.1, the ratio $\Omega/n$ governs the average value of the torque exerted on each DNA bead, so that the arched stripe and counter-arch patterns were observed only above a threshold that lies somewhere between $\Omega/n = 500$ rad×s$^{-1}$×bead$^{-1}$ and $\Omega/n = 1000$ rad×s$^{-1}$×bead$^{-1}$, for which the torque is larger than the buckling threshold. Simulations furthermore indicate that, above the threshold, the growth rate of the arched stripe and counter-arch patterns is nearly proportional to $\Omega$. This is clearly seen in the seventh column of Table 1, as well as in Figure 7, where the blue dashed line shows the time evolution of $k_{\mathrm{apex}}$ for a simulation with $\Omega/n = 1000$ rad×s$^{-1}$×bead$^{-1}$. Its slope is approximately one half of the slope of the brown and red lines, which were obtained from simulations with $\Omega/n = 2000$ rad×s$^{-1}$×bead$^{-1}$. Figure A2 shows superposed contact maps, the red one being computed at $t = 20$ ms from a simulation with $\Omega/n = 2000$ rad×s$^{-1}$×bead$^{-1}$ (same map as in Figure 3), while the green one is computed at $t = 40$ ms from a simulation with $\Omega/n = 1000$ rad×s$^{-1}$×bead$^{-1}$. It is seen that doubling the waiting time for the second simulation indeed ensures that the contact patterns in Figure A2 have nearly the same size, in spite of the fact that $\Omega$ for the second simulation is only one half of its value for the first simulation. Still, one may notice that the contact patterns are not completely identical, in the sense that a smaller value of $\Omega$ clearly favors a comparatively longer $(b)$ contact line (edges of the arched stripe), at the expense of comparatively shorter $(p)$ and $(a)$ contact lines (central part of the arched stripe and counter-arch). This observation is confirmed by the comparison of the last two columns of Table 1.

*Appendix A.3. Influence of the length of the transcription track*

Simulations with a 450 beads transcription track were performed to check how the length of the transcription track impacts the contact pattern. The outcome of these simulations is somewhat counter-intuitive, in the sense that the contact pattern grows more slowly for a 450 beads track than for a 150 beads one, whereas one could naively have expected the opposite to be true. This is illustrated in Figure 7, where the green dot-dashed line shows the time evolution of $k_{\mathrm{apex}}$ for a simulation with a 450 beads track and $\Omega/n = 2000$ rad×s-1×bead-1. It is seen that its slope is significantly smaller than that of the red solid curve, which was obtained with a 150 beads track. This point is also confirmed by Figure A3, which shows superposed contact maps computed at $t = 25$ ms from simulations with $\sigma = -6.3$ %, $\Omega/n = 2000$ rad×s-1×bead-1 and a 150 beads track (red map) or a 450 beads one (green map). This figure indicates that the $(P^-)$ plectoneme and the $(p)$ contact line are indeed significantly shorter for the 450 beads track than for the 150 beads one. In contrast, the $(B^+)$ and $(B^-)$ segments and the $(b)$ contact line are as long for the 450 beads track as for the 150 beads one. The main reason, why translocation along a 450 beads track results in a shorter $(P^-)$ plectoneme is that the polymerase moves too fast and over too long a distance for $(P^-)$ to adapt to its motion [18]. Actually, $(P^-)$ detaches from the upstream side of the polymerase somewhere near the middle of the track, so that further translocation of the polymerase till the end of the track no longer contributes to increase the size of $(P^-)$, as is the case in Figure 8. $(P^-)$ does reattach to the polymerase and its length increases again when the polymerase starts a new translocation from the beginning of the track, but this intermittent process is less efficient than the continuous process that takes place for the 150 beads translocation track.

| Track length (beads) | $\sigma$ (%) | $\Omega/n$ (rad×s$^{-1}$×bead$^{-1}$) | Pattern (Y/N) | Location (U/U+D) | Stability (Y/N) | $(P^-)$ growth rate (beads/ms) | $(B^+)$ growth rate (beads/ms) |
|---|---|---|---|---|---|---|---|
| 150 | 0.0 | 250 | N | | | | |
| | | 500 | N | | | | |
| | | 1000 | Y | U+D | N | | |
| | | 2000 | Y | U+D | N | | |
| | -0.8 | 2000 | Y | U+D | N | | |
| | -1.6 | 2000 | Y | U | N | | |
| | -2.4 | 2000 | Y | U | Y | 63 | 43 |
| | -3.2 | 250 | N | | | | |
| | | 500 | N | | | | |
| | | 1000 | Y | U | Y | 36 | 40 |
| | | 2000 | Y | U | Y | 57 | 55 |
| | -6.3 | 250 | N | | | | |
| | | 500 | N | | | | |
| | | 1000 | Y | U | Y | 33 | 46 |
| | | 2000 | Y | U | Y | 61 | 65 |
| 450 | -6.3 | 250 | N | | | | |
| | | 500 | N | | | | |
| | | 1000 | Y | U | N | | |
| | | 2000 | Y | U | N | | |

**Table 1** : **Summary of explored parameter combinations**. For each set of parameters (track length, $\sigma$ and $\Omega/n$), column "Pattern" indicates whether the contact pattern consisting of arched stripe, bundled domain and counter-arch is observed (Y) or not (N); column "Location" whether the arched stripe forms only upstream of the polymerase (U) or alternately on both sides of the polymerase (U+D); column "Stability" whether the pattern remains stable till it nearly fills the DNA chain (Y) or whether plectoneme $(P^-)$ detaches much sooner from the polymerase (N). Finally, the last two columns indicate, for stable patterns, the growth rate of the $(P^-)$ plectoneme and the $(B^+)$ segment.

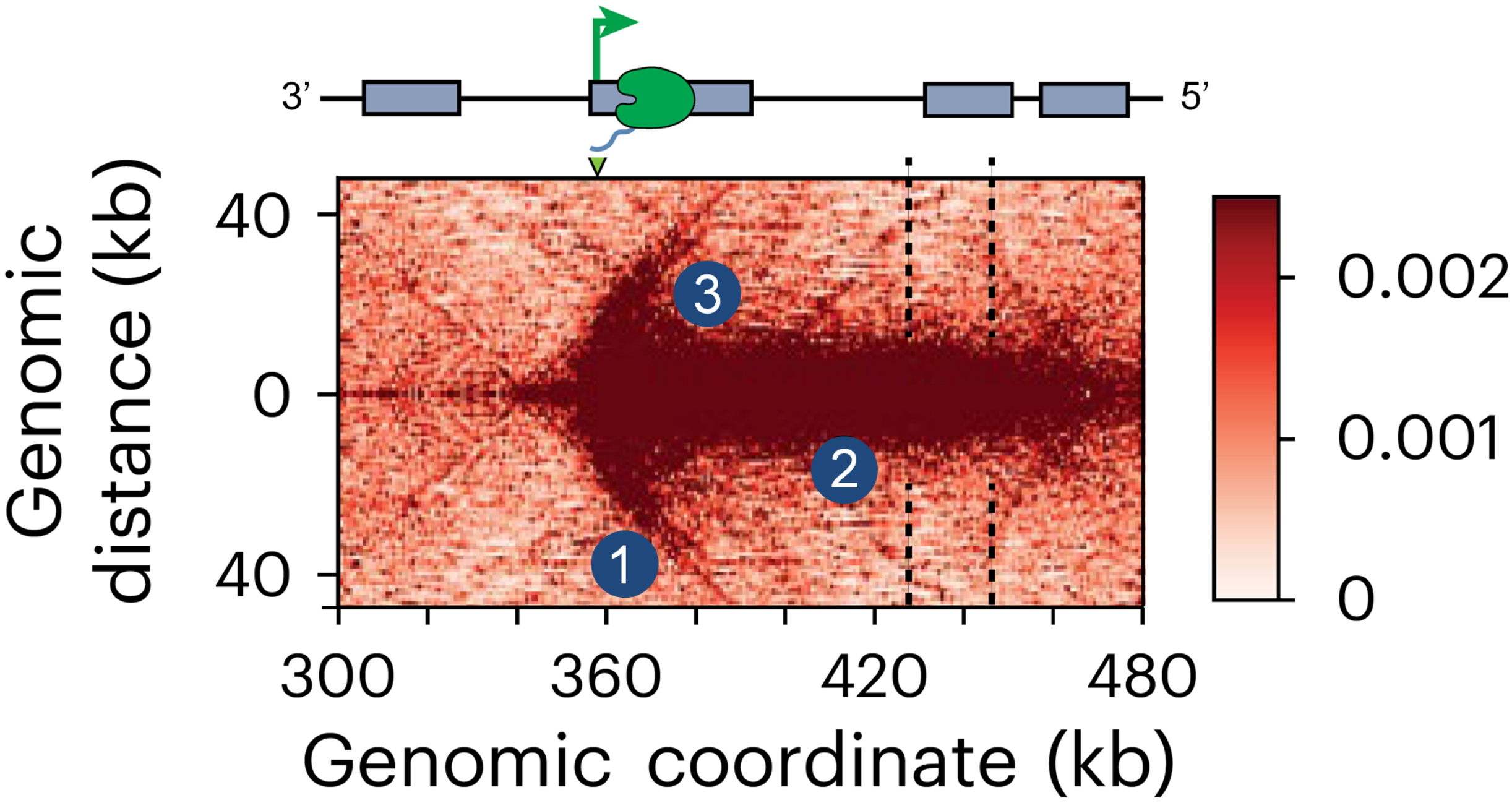

**Figure 1** : **Experimental Hi-C contact map**. Magnification of the T7 promoter region in the normalized contact map of a single, active transcription unit within the entire E. coli genome. T7 promoter is turned on, while all endogenous genes are turned off by rifampicin. A schematic representation of the region's genetic content is presented on the top and the locations of silenced endogeneous secDEF and cyoABCD operons are shown as vertical dashed lines in the contact map. The numbered labels on the map highlight the features discussed in the text: (1) arched stripe pattern; (2) bundled domain; (3) counter-arch (reproduced and adapted from Figure 2(b) of [15]).

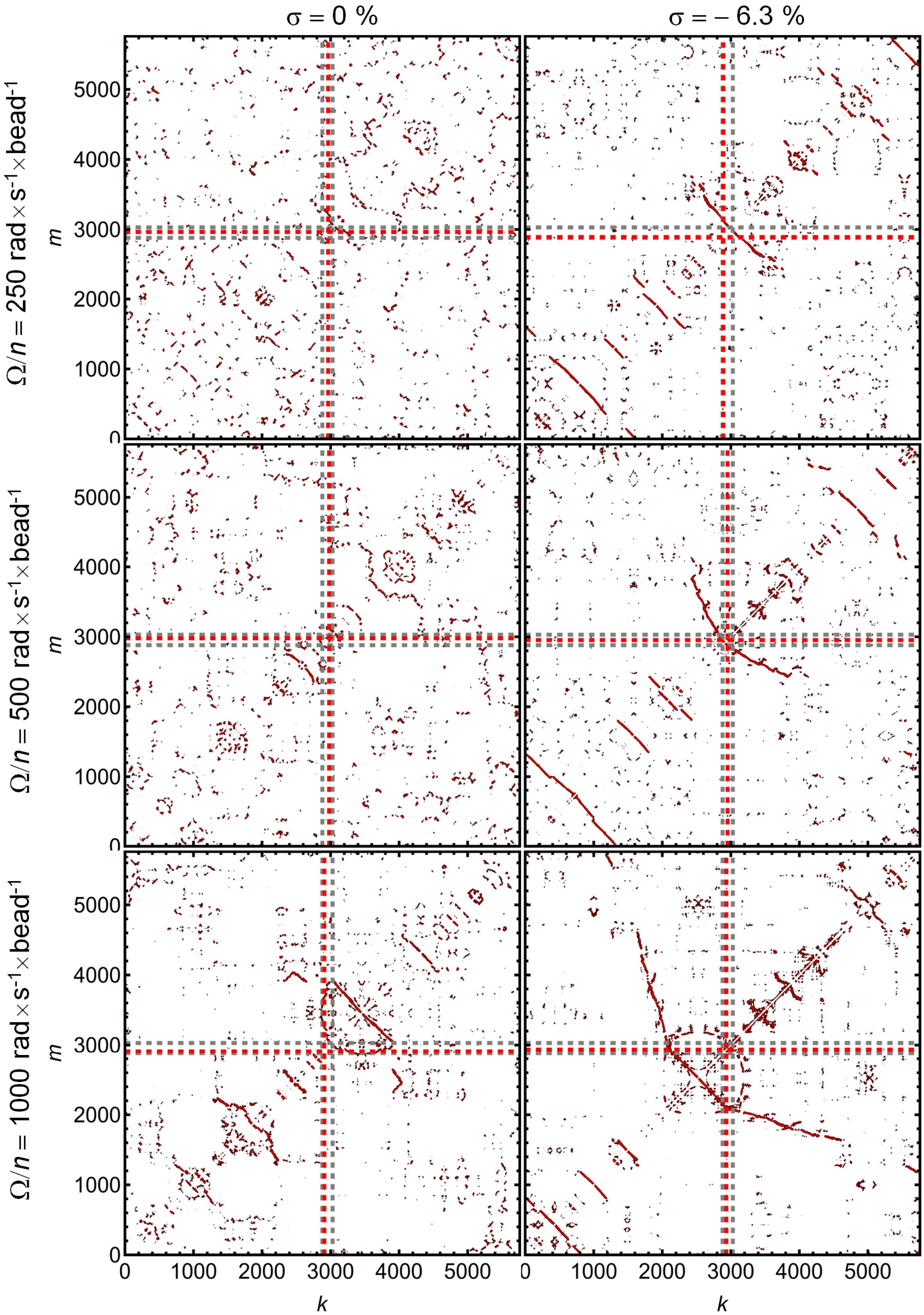

**Figure 2** : **Threshold at $\Omega/n$ =1000 rad×s$^{-1}$×bead$^{-1}$**. Representative contact maps computed at $t = 40$ ms from simulations involving a polymerase that repeatedly translocates along a 150 beads DNA track, with $\sigma = 0$ (left) or $\sigma = -6.3$ % (right), and $\Omega/n$ =250 (top), 500 (middle) or 1000 (bottom) rad×s$^{-1}$×bead$^{-1}$. $k$ and $m$ are the indexes of DNA beads. The horizontal and vertical gray dashed lines indicate the positions of the beginning and the end of the transcription track. The red dashed lines indicate the position of the polymerase.

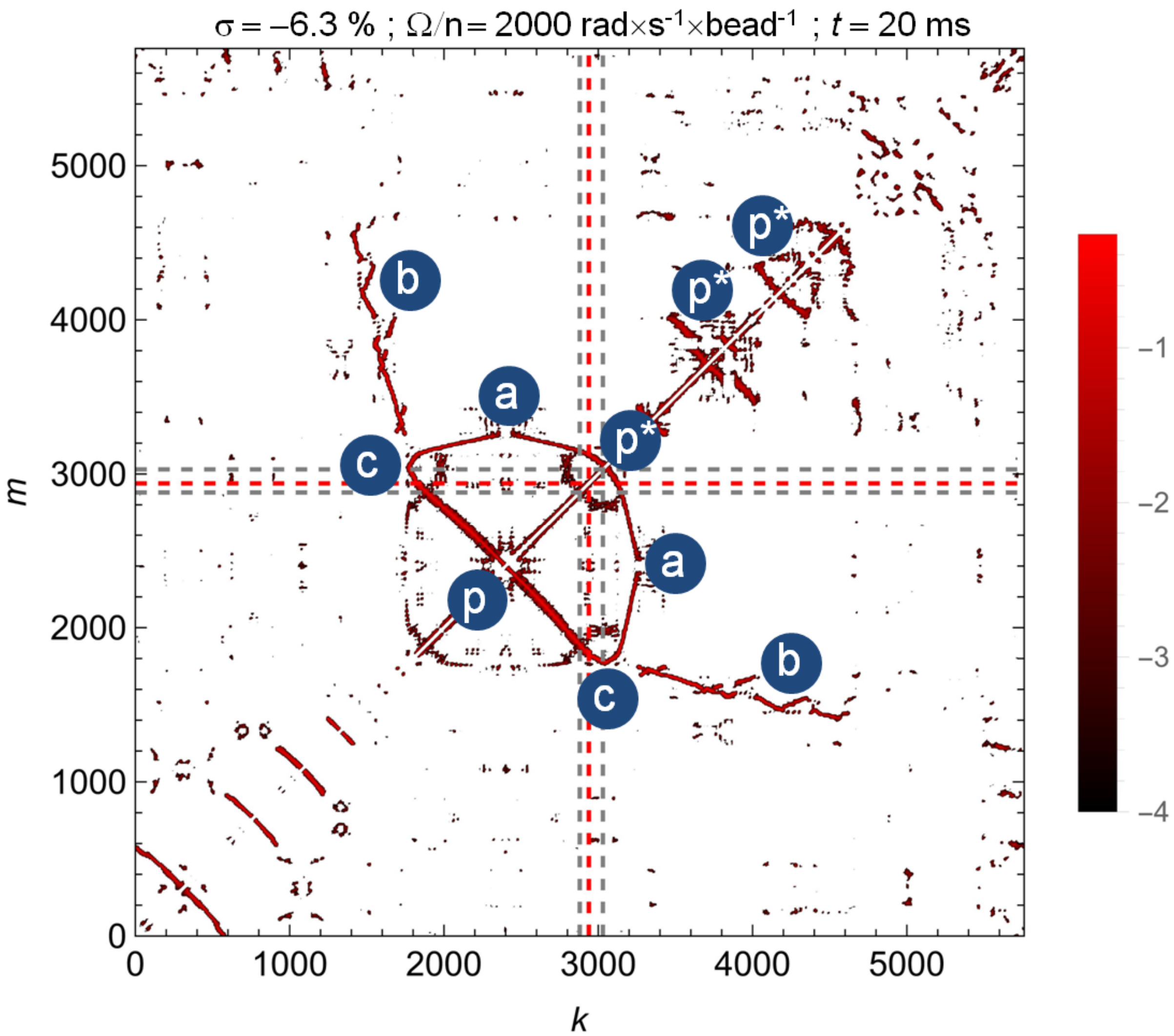

**Figure 3** : **Arched stripe pattern in BD simulations**. Contact map computed at $t = 20$ ms from a simulation with $\sigma = -6.3\,\%$ and $\Omega/n = 2000$ rad×s$^{-1}$×bead$^{-1}$. $k$ and $m$ are the indexes of DNA beads. The horizontal and vertical gray dashed lines indicate the positions of the beginning and the end of the transcription track. The red dashed line indicates the position of the polymerase.

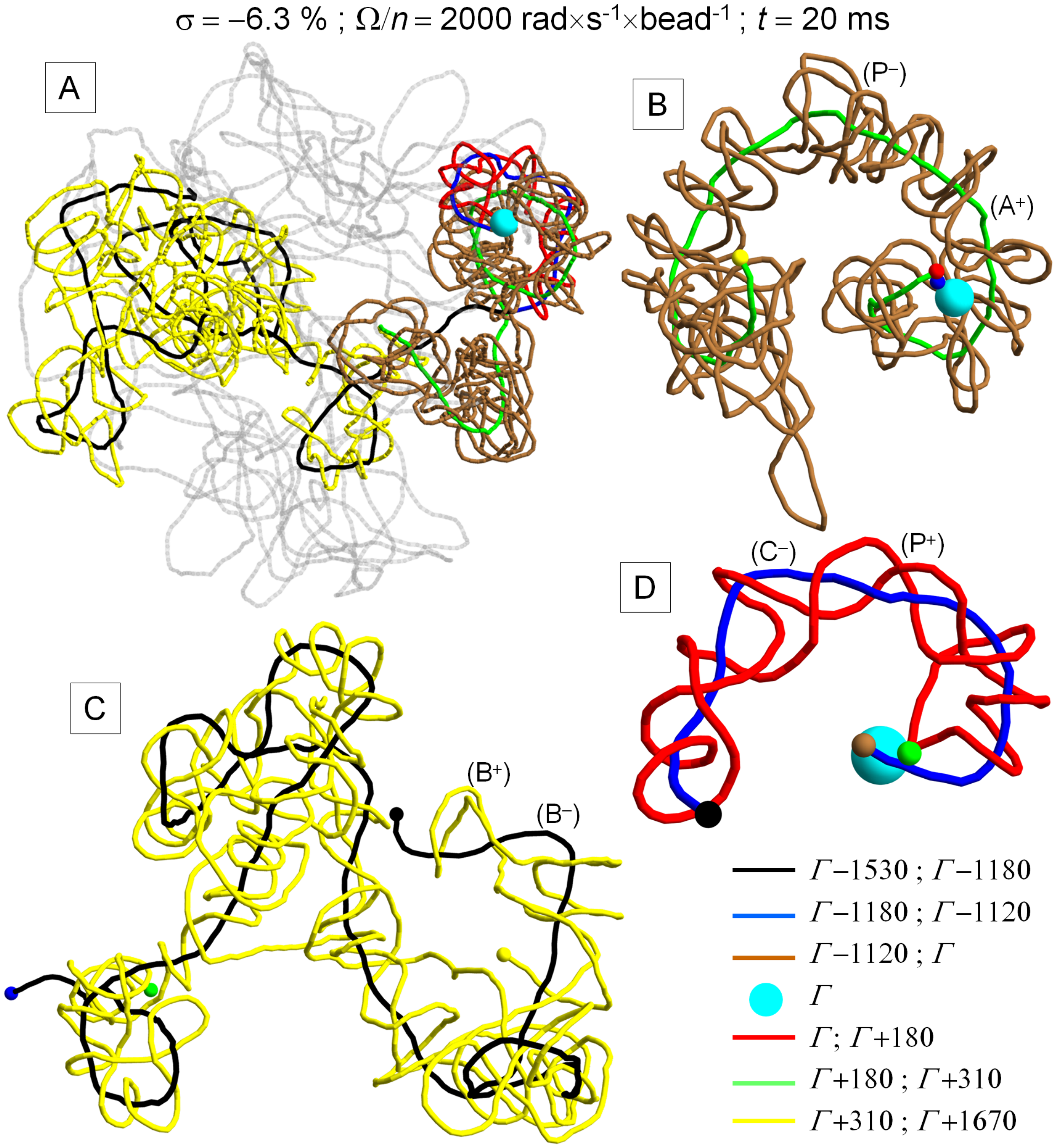

**Figure 4** : **Conformation of the DNA coil corresponding to the contact map in Figure 3**. This snapshot is extracted at $t = 20$ ms from a simulation with $\sigma = -6.3$ % and $\Omega/n = 2000$ rad×s-1×bead-1. Panel A shows the complete coil. DNA segments involved in contacts lines (a)-(c), (p) and (p*) of Figure 3 are shown as colored tubes, while the rest of the DNA chain is shown as a gray transparent tube. Panels B, C and D zoom in on the specific regions that are responsible for these contact lines. The legend indicates the position of each segment with respect to the position $\Gamma$ of the polymerase.

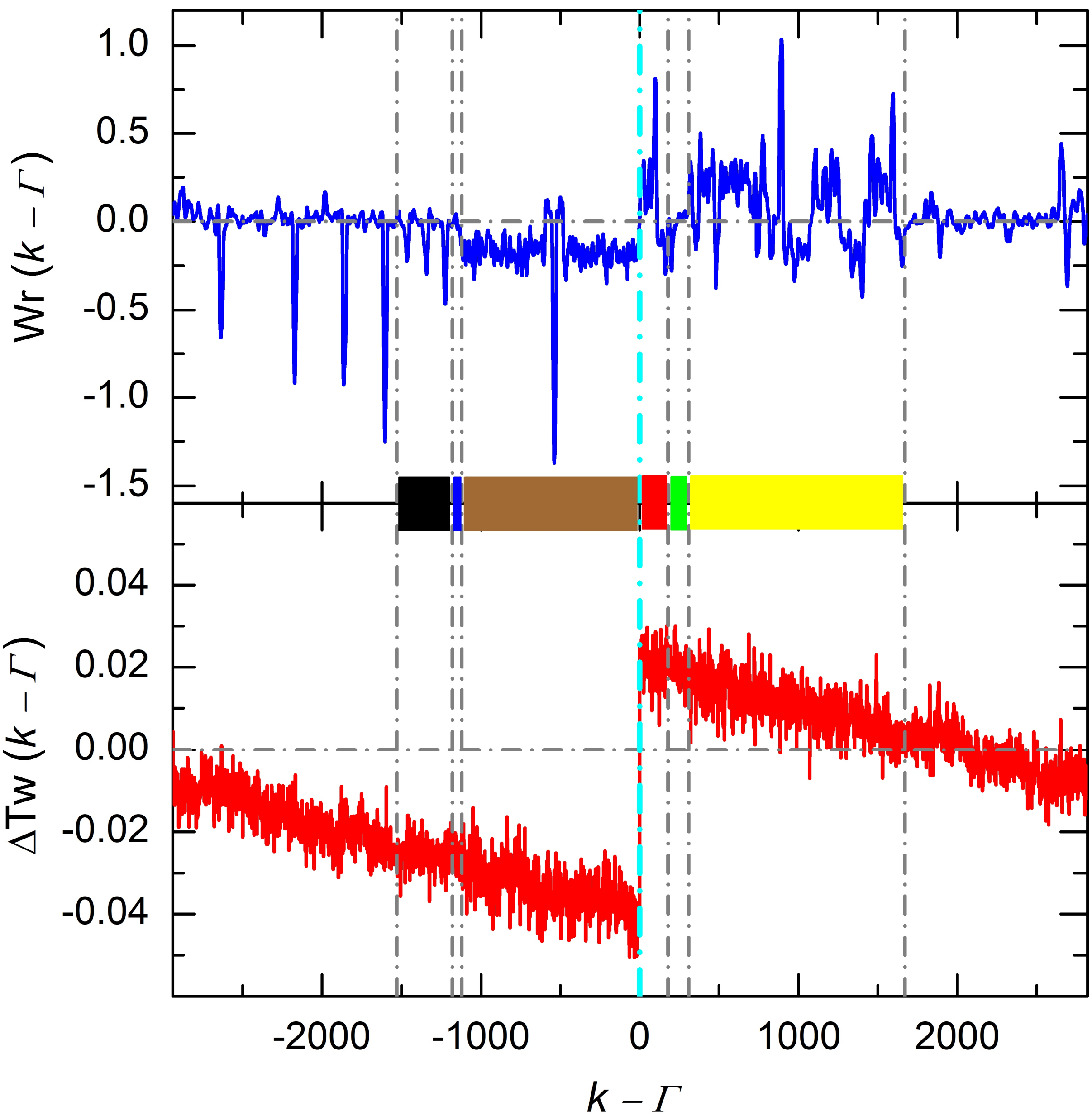

**Figure 5 : Plot of $Wr(k-\Gamma)$ and $\Delta Tw(k-\Gamma)$ for the DNA conformation in Figure 4.** $\Gamma$ denotes the position of the polymerase, $Wr(k)$ the local writhe at bead $k$ and $\Delta Tw(k)$ the local excess of twist at bead $k$. This conformation is extracted at $t = 20$ ms from a simulation with $\sigma = -6.3\,\%$ and $\Omega/n = 2000$ rad×s$^{-1}$×bead$^{-1}$. The vertical gray dot-dashed lines indicate the limits of the DNA segments shown in color in Figure 4. The corresponding colors are reproduced on the central horizontal axis, for the sake of an easier comparison with Figure 4.

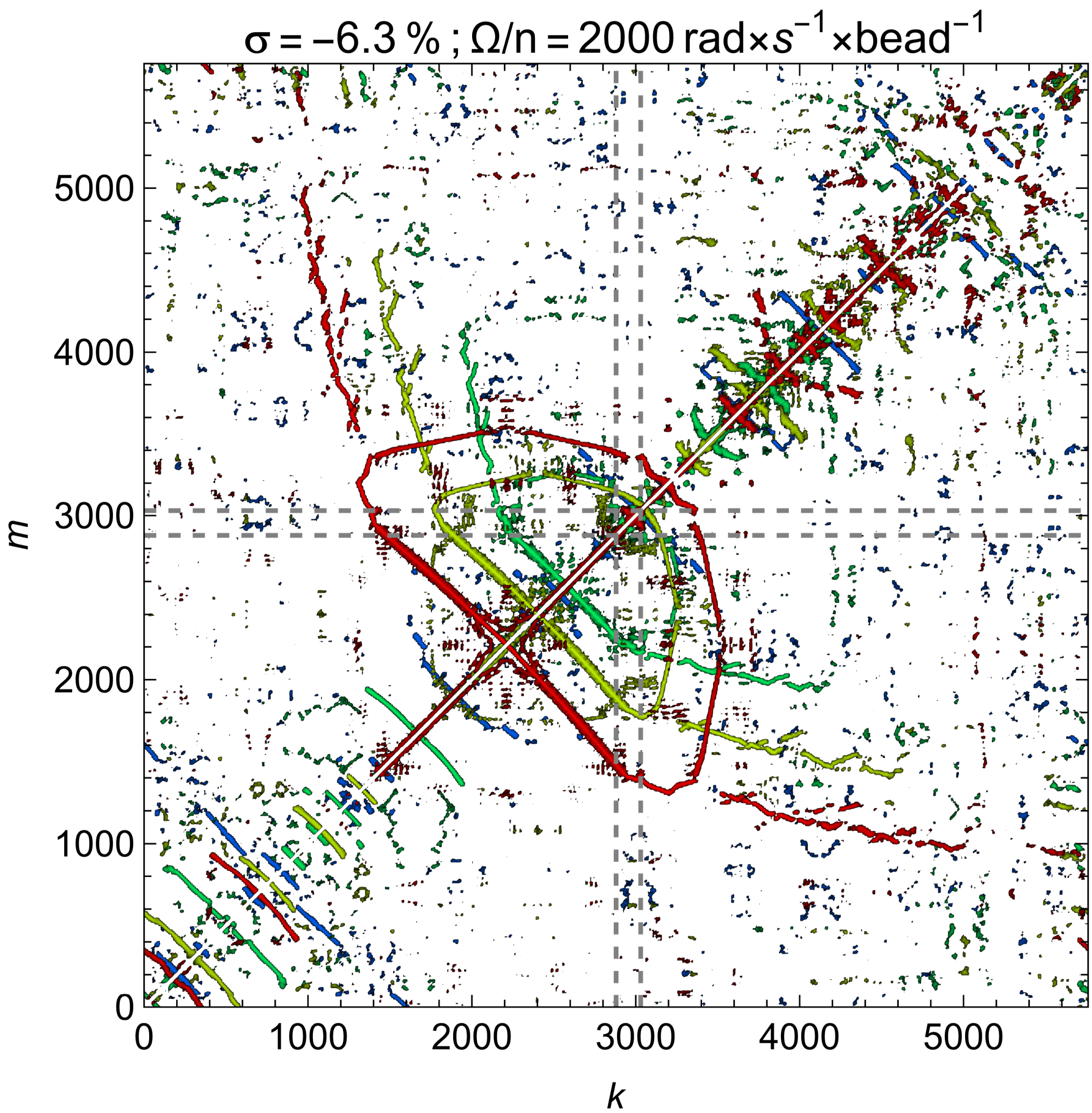

**Figure 6** : **Linear expansion of the arched stripe**. Superposed contact maps computed at $t = 0$ (blue), 10 (green), 20 (yellowish green) and 30 (red) ms from a simulation with $\sigma = -6.3\ \%$ and $\Omega/n = 2000$ rad×s$^{-1}$×bead$^{-1}$. $k$ and $m$ are the indexes of DNA beads. The horizontal and vertical gray dashed lines indicate the positions of the beginning and the end of the transcription track.

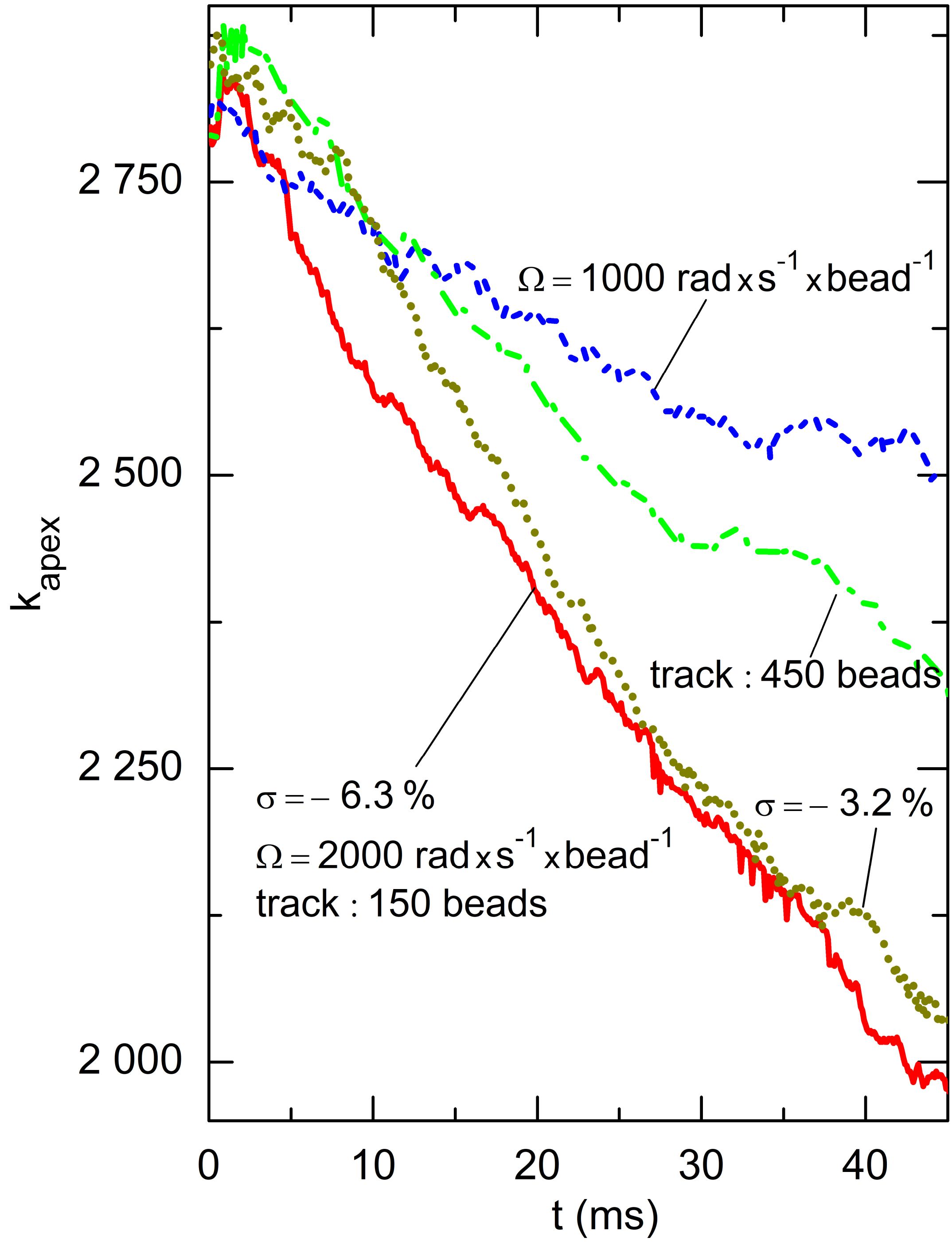

**Figure 7** : **Time evolution of $\boldsymbol{k}_{\mathbf{apex}}$**. $k_{\text{apex}}$ is the position of the apex of the $(P^{-})$ plectoneme. The red solid line is obtained from a simulation with $\sigma = -6.3$ %, $\Omega/n = 2000$ rad×s-1×bead-1 and a 150 beads DNA track, the brown dotted line from a simulation with $\sigma = -3.2$ %, $\Omega/n = 2000$ rad×s-1×bead-1 and a 150 beads DNA track, the green dot-dashed line fom a simulation with $\sigma = -6.3$ %, $\Omega/n = 2000$ rad×s-1×bead-1 and a 450 beads DNA track, and the blue dashed line from a simulation with $\sigma = -6.3$ %, $\Omega/n = 1000$ rad×s-1×bead-1 and a 150 beads DNA track.

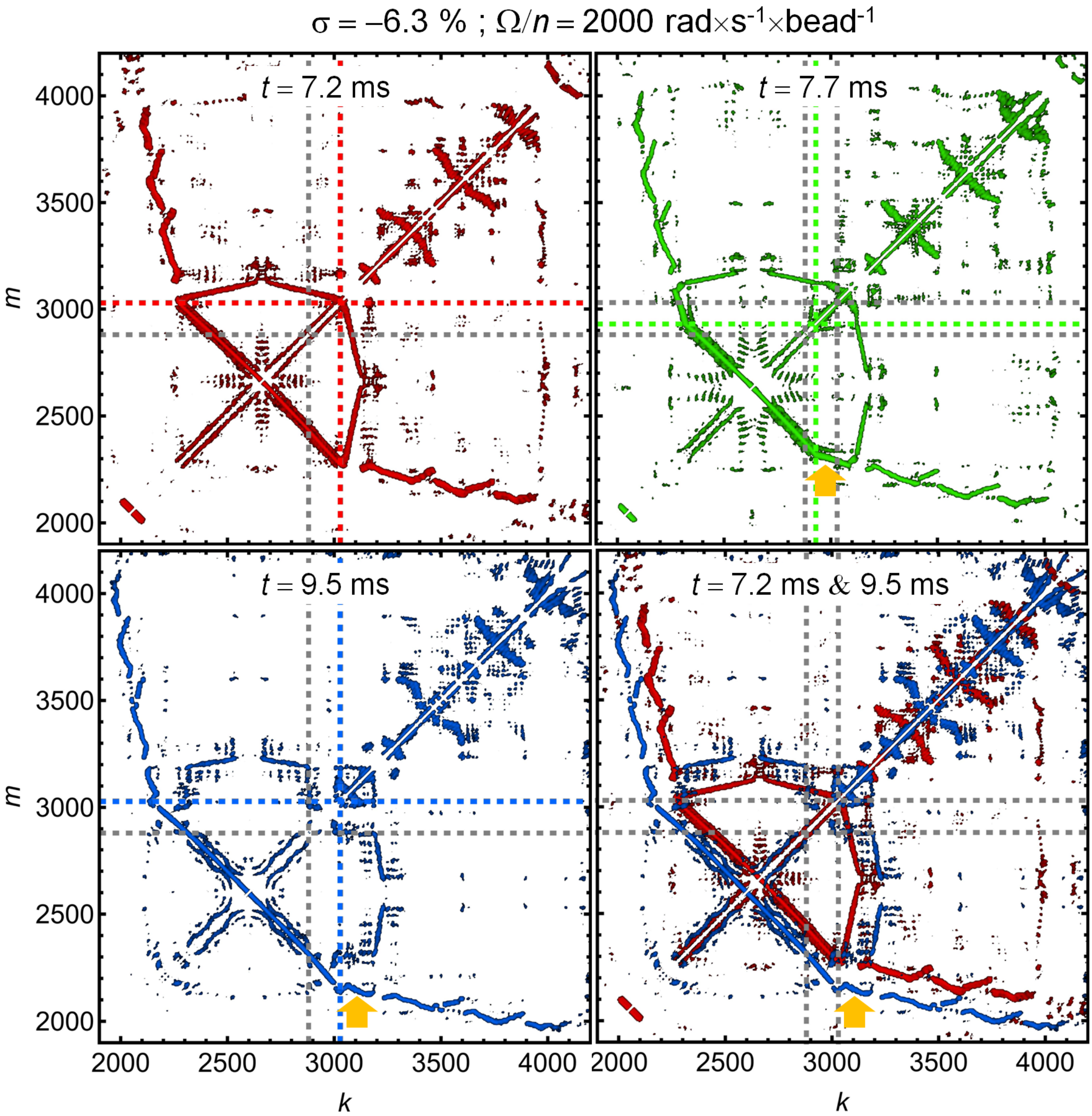

**Figure 8** : **Expansion dynamics of the arched stripe**. Contact maps computed at $t = 7.2$ ms (top left panel), 7.7 ms (top right panel) and 9.5 ms (bottom left panel) from a simulation with $\sigma = -6.3\,\%$ and $\Omega/n = 2000$ rad×s$^{-1}$×bead$^{-1}$. The bottom right panel shows the superposition of the contact maps at $t = 7.2$ ms and 9.5 ms. $k$ and $m$ are the indexes of DNA beads. The horizontal and vertical gray dashed lines indicate the positions of the beginning and the end of the transcription track. The orange arrow shows the region of the contact map, which is affected by the destabilization of one foot of the $(P^{-})$ plectoneme and its transfer to the $(B^{-})$ segment.

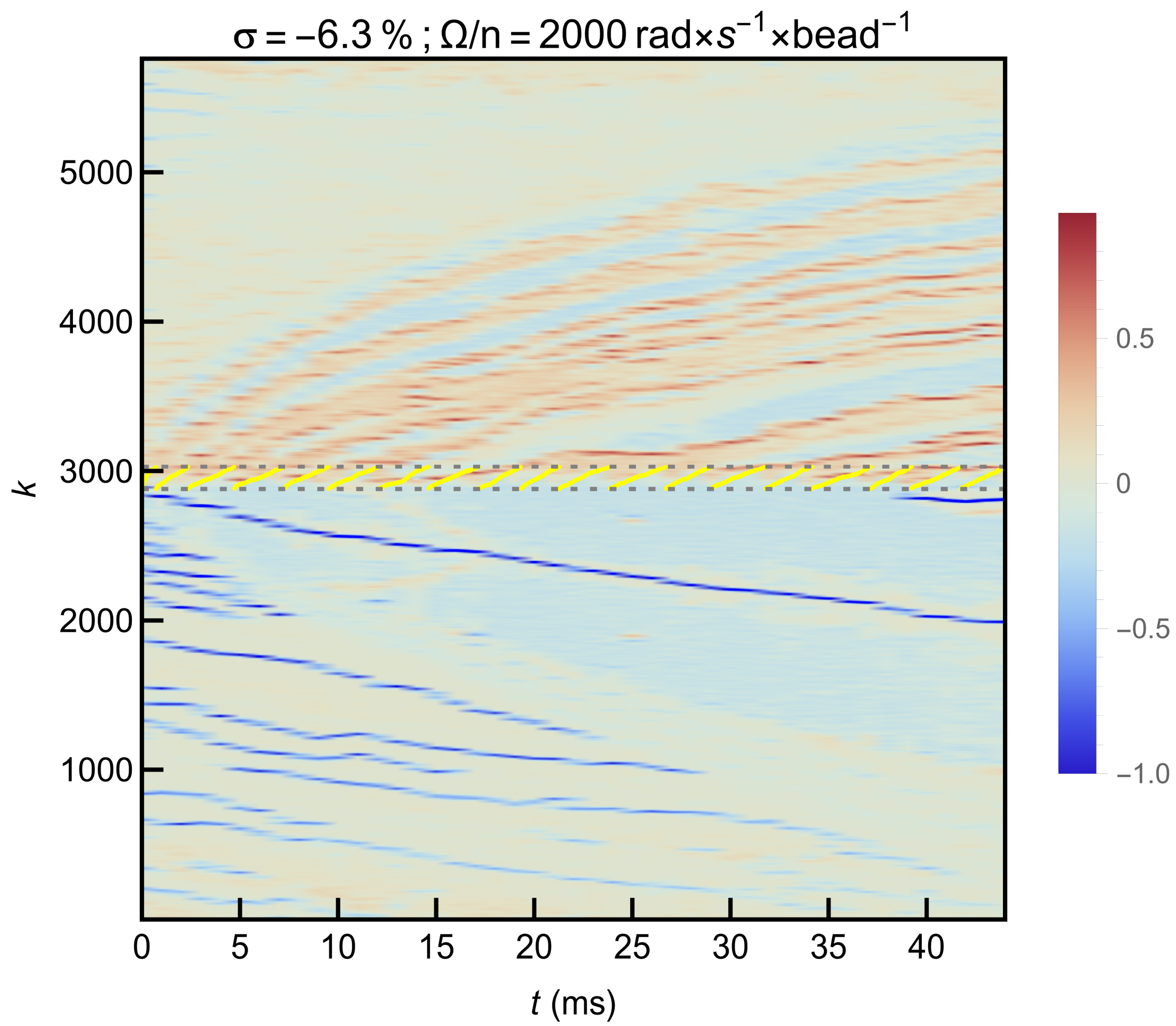

**Figure 9** : **Time evolution of $\boldsymbol{Wr}(\boldsymbol{k})$.** This plot is obtained from a simulation with $\sigma = -6.3$ % and $\Omega/n = 2000$ rad×s$^{-1}$×bead$^{-1}$. The values of the local writhe $Wr(k)$ are represented according to a color code, which ranges from deep blue for $Wr(k) = -1$ to deep red for $Wr(k) = 1$. The horizontal gray dashed lines indicate the positions of the beginning and the end of the transcription track. The yellow line represents the position of the polymerase.

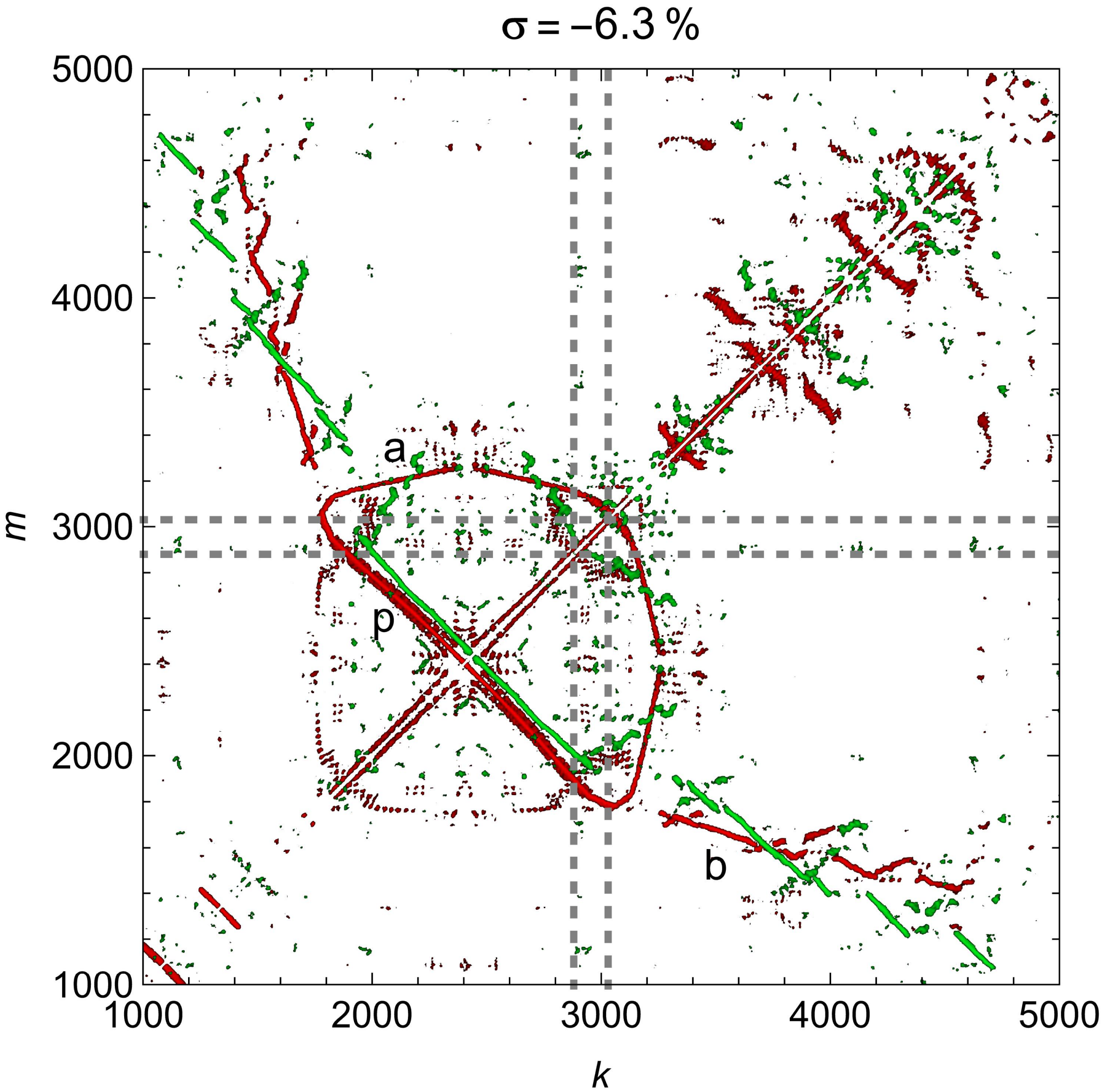

**Figure 10** : **Toroidal supercoiling is an out-of-equilibrium feature**. This figure displays two superposed contact maps. The red map is computed at $t = 20$ ms from a simulation with $\sigma = -6.3$ % and $\Omega/n = 2000$ rad×s-1×bead-1 (same map as in Figure 3) The green map is obtained from the red one by letting the system evolve for an additional 6 ms in the absence of polymerase translocation, that is with $\Omega/n = 0$ rad×s-1×bead-1. The horizontal and vertical gray dashed lines indicate the positions of the beginning and the end of the transcription track.

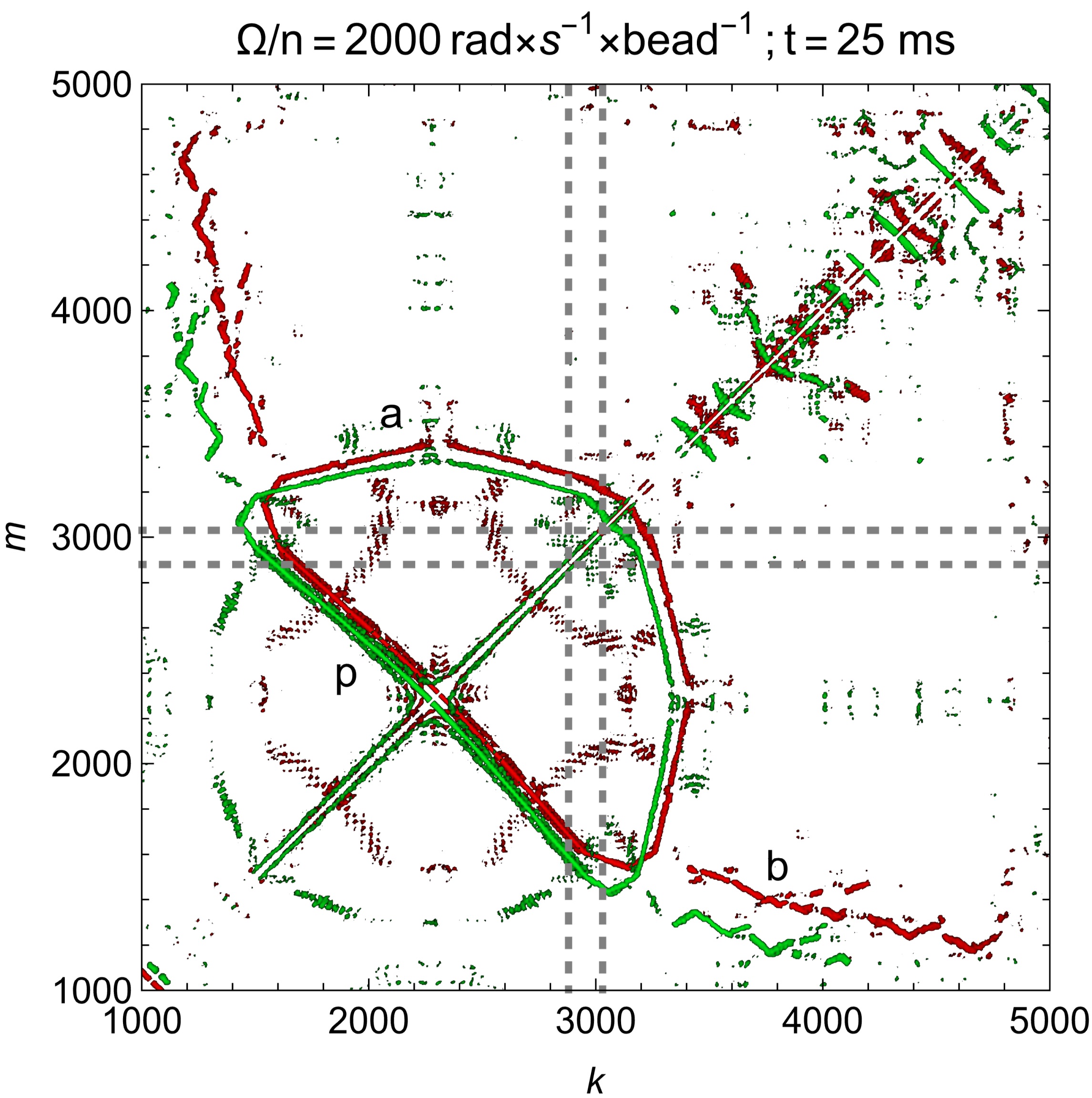

**Figure A1** : **Influence of $\sigma$**. Superposed contact maps computed at $t = 25$ ms from simulations with $\Omega/n = 2000$ rad×s$^{-1}$×bead$^{-1}$ and $\sigma = -6.3$ % (red map) or $\sigma = -2.4$ % (green map). The horizontal and vertical gray dashed lines indicate the positions of the beginning and the end of the transcription track.

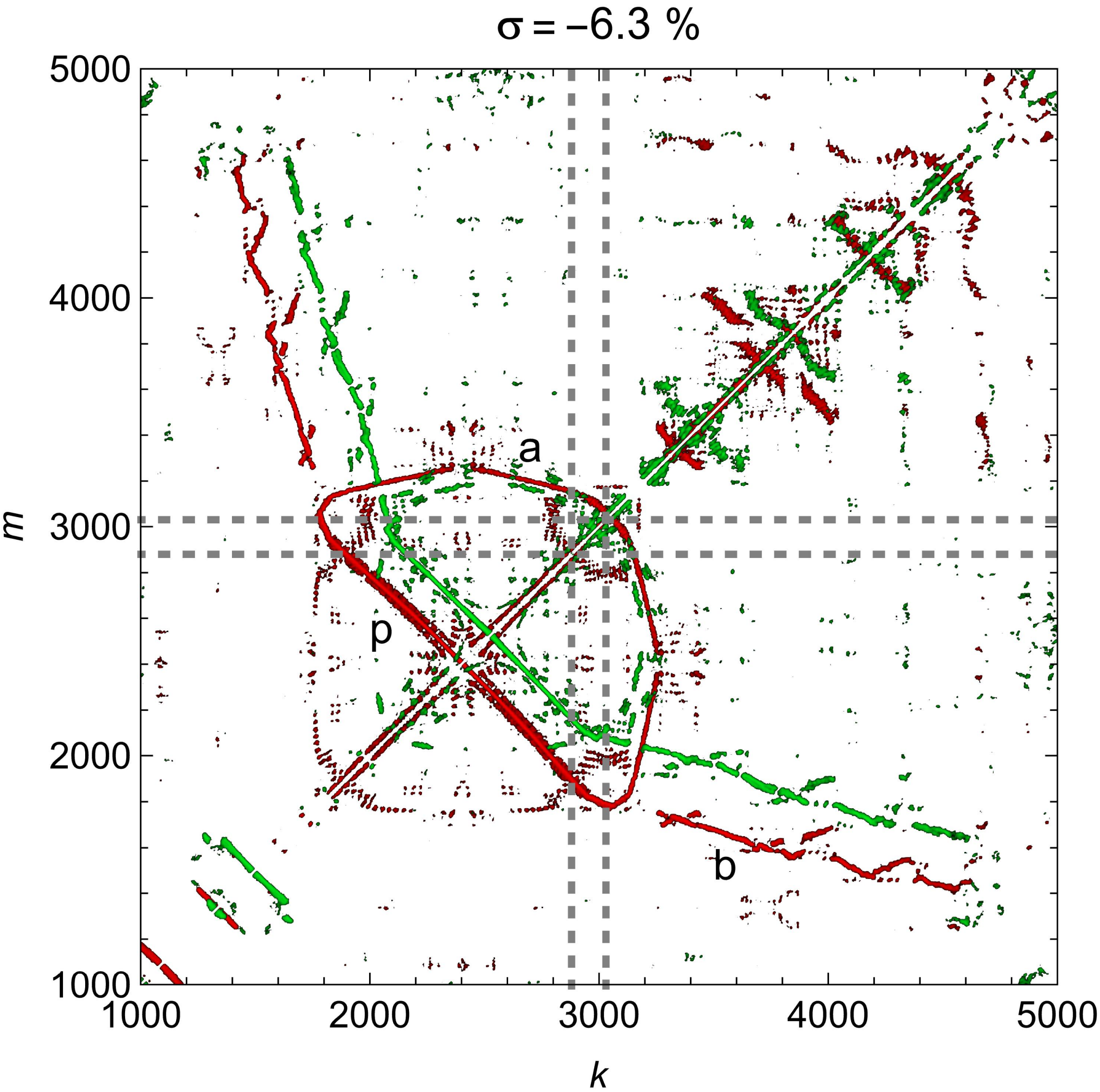

**Figure A2** : **Influence of $\Omega/n$**. This figure displays two superposed contact maps. The red map is computed at $t = 20$ ms from a simulation with $\sigma = -6.3$ % and $\Omega/n = 2000$ rad×s$^{-1}$×bead$^{-1}$ (same map as in Figure 3). The green map is computed at $t = 40$ ms from a simulation with $\sigma = -6.3$ % and $\Omega/n = 1000$ rad×s$^{-1}$×bead$^{-1}$. The horizontal and vertical gray dashed lines indicate the positions of the beginning and the end of the transcription track.

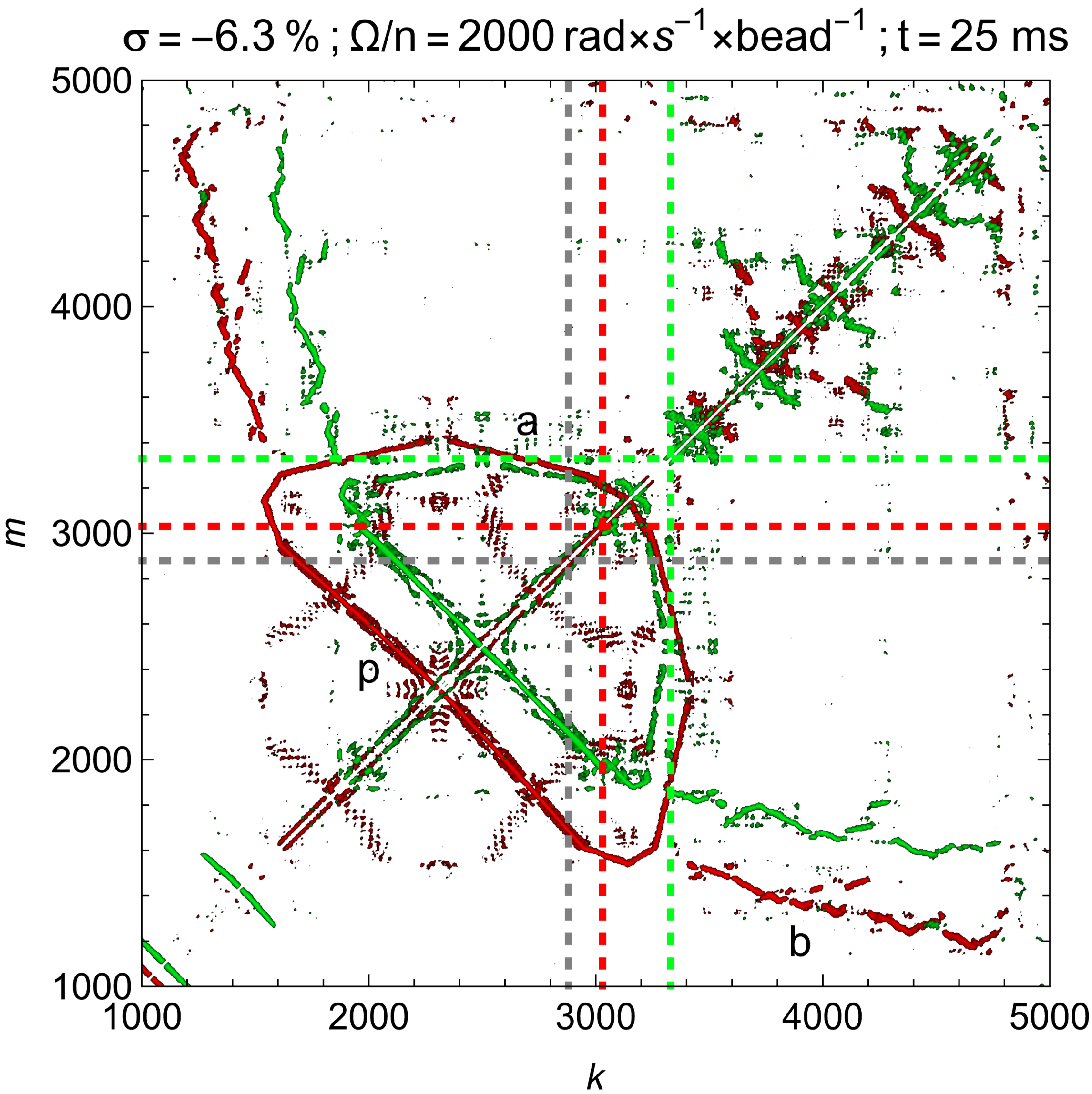

**Figure A3** : **Influence of the length of the transcription track**. Superposed contact maps computed at $t = 25$ ms from simulations with $\sigma = -6.3$ %, $\Omega/n = 2000$ rad×s$^{-1}$×bead$^{-1}$ and a 150 beads track (red map) or a 450 beads one (green map). The horizontal and vertical dashed lines indicate the positions of the beginning and the end of the transcription tracks. Both tracks start at the gray dashed line. The 150 beads track ends at the red dashed line, while the 450 beads track ends at the green dashed line.